\documentclass[letterpaper,english,aps,preprint,superscriptaddress,
nofootinbib,byrevtex]{revtex4}
\usepackage[T1]{fontenc}
\usepackage[latin1]{inputenc}
\usepackage{subfigure}
\usepackage{amsmath}
\usepackage{graphicx}
\usepackage{amssymb}

\makeatletter

\newcommand{\etal}{\textit{et al.}}

\newcommand{\sst}[1]{\scriptstyle{#1}}
\newcommand{\ssst}[1]{\scriptscriptstyle{#1}}

\newcommand{\bra}[1]{\langle #1 \vert}
\newcommand{\ket}[1]{\vert #1 \rangle}
\newcommand{\expec}[1]{\langle #1 \rangle}

\newcommand{\mN}{m_{\ssst{N}}}
\newcommand{\sstw}{\sin^2\theta_W}

\newcommand{\GMs}{G_{\ssst{M}}^{(s)}}
\newcommand{\GAeV}{G_{\ssst{A}}^{(e)\,\sst{(T=1)}}}
\newcommand{\gAe}{g_{\ssst{A}}^{e}}
\newcommand{\gVe}{g_{\ssst{V}}^{e}}

\newcommand{\WEM}{W^{\ssst{EM}}}
\newcommand{\WPV}{W^{\ssst{PV}}}
\newcommand{\WPVZ}{W^{\ssst{PV(Z)}}}
\newcommand{\WPVg}{W^{\ssst{PV}(\gamma)}}
\newcommand{\HPV}{H_{\ssst{PV}}}
\newcommand{\ALR}{A_{\ssst{LR}}}
\newcommand{\ALRZ}{A_{\ssst{LR}}^{(Z)}}
\newcommand{\ALRg}{A_{\ssst{LR}}^{(\gamma)}}

\newcommand{\vect}{\vec{\tau}}
\newcommand{\vecs}{\vec{\sigma}}

\usepackage{graphics}
\usepackage{subfigure}

\usepackage{babel}
\makeatother
\begin{document}

\title{Hadronic Parity Violation and Inelastic Electron-Deuteron Scattering}

\author{C.-P. Liu}

\affiliation{TRIUMF, 4004 Wesbrook Mall, Vancouver, B.C., Canada V6T 2A3}

\email{cpliu@triumf.ca}

\author{G. Pr\'{e}zeau}

\affiliation{Kellogg Radiation Laboratory, Caltech, Pasadena, CA 91125, USA}

\email{prezeau@krl.caltech.edu}

\author{M.J. Ramsey-Musolf}

\affiliation{Kellogg Radiation Laboratory, Caltech, Pasadena, CA 91125, USA}

\affiliation{Department of Physics, University of Connecticut, Storrs, 
CT 06269,
USA}

\affiliation{Institute for Nuclear Theory, Box 351550, University of 
Washington, Seattle, WA 98195-1550, USA}

\email{mjrm@krl.caltech.edu}

\begin{abstract}
We compute contributions to the parity-violating (PV) inelastic
electron-deuteron scattering asymmetry arising from hadronic PV. While
hadronic PV effects can be relatively important in PV threshold electro-
disintegration,
we find that they are highly suppressed at quasielastic kinematics.
The interpretation of the PV quasielastic asymmetry is, thus, largely
unaffected by hadronic PV.
\end{abstract}
\maketitle

\section{Introduction \label{sec:intro}}

Parity-violating (PV) inelastic electron-nucleus scattering
is an important tool in the study of hadron structure \cite{Mu..94}.
In combination with PV elastic electron-proton ($e$-$p$) scattering,
measurements of the PV quasielastic (QE) electron-deuteron ($e$-$d$) asymmetry
allow a separate determination of the strangeness magnetic form factor,
$\GMs (Q^{2})$, and the isovector axial vector form factor, $\GAeV (Q^{2})$.
Knowledge of $\GMs (Q^{2})$ provides a window on the role played
by sea-quarks in the electromagnetic structure of the nucleon. The
axial vector form factor, in contrast, is sensitive to nucleon structure
effects in higher-order, electroweak radiative corrections. These
corrections, which depend on the species of lepton probe (hence, the
{}``$e$'' superscript), share features with corrections relevant
to other precision electroweak measurements, such as the PV asymmetry
in polarized neutron $\beta $-decay. The proper interpretation of
such measurements relies on an adequate understanding of electroweak
radiative corrections \cite{McMu02}.

Recently, the SAMPLE collaboration has performed a separate determination
of $\GMs $ and $\GAeV $ at $Q^{2}=0.1$ (GeV/$c$)$^{2}$ using
PV $e$-$p$ and PV QE $e$-$d$ scattering \cite{SAMPLE97,SAMPLE00a,SAMPLE00b}.
The results indicate a value for $\GAeV $ consistent with zero. At
tree-level, one expects $\GAeV (Q^{2}=0)=-1.267$, while radiative
corrections reduce the magnitude by roughly $40\pm 20\%$ \cite{Zh..00,MuHo90}.
These corrections include potentially significant hadronic contributions
that are responsible for the estimated theoretical uncertainty. To
make the measured value of $\GAeV $ close to zero would require additional
effects not included in the calculation of Refs. \cite{Zh..00,MuHo90}.

One possibility, which we explore in this paper, is the contribution
from the PV nucleon-nucleon ($NN$) interaction. The latter induces
small admixtures of opposite parity states into the deuteron as well
as the scattered $np$ partial waves. These parity admixtures contribute
to the PV asymmetry when a $\gamma $ is exchanged between the electron
and target%
\footnote{Current conservation also implies the presence of PV electromagnetic
two-body current contributions, as we discuss below.%
}. Moreover, in contrast to the effect of $Z^{0}$ exchange, these
hadronic PV effects in $\gamma $-exchange give rise to a term in
the asymmetry which does not vanish at $Q^{2}$ = 0. For sufficiently
small $Q^{2}$, this term would dominate the asymmetry. One might
ask, then, whether the omission of this term in the interpretation
of the SAMPLE deuterium measurement is responsible for the apparent,
anomalously large radiative corrections entering $\GAeV $.

Below, we show that the magnitude of this $Q^{2}$-independent hadronic
PV contribution is too small to account for the observed $\GAeV $
effect. Based on simple scaling arguments, the relative importance
of the $Q^{2}$-independent contribution -- compared to the {}``canonical''
$Z^{0}$-exchange induced asymmetry -- goes as $\sim 10^{-4}\mN ^{2}/Q^{2}$.
Thus, at the SAMPLE kinematics, $Q^{2}=0.1$ (GeV/$c$)$^{2}$, we
expect the hadronic PV contribution to generate at most a few parts
in a thousand correction to the asymmetry -- far short of what would
be needed to close the gap between the theoretical and experimental
values for $\GAeV $.

We also carry out an explicit calculation of the hadronic PV contribution
and verify the expectations based on these scaling arguments. Our
computation follows on the work of Refs. \cite{HwHe80,HwHM81}, in
which the hadronic PV contribution to PV threshold deuteron electro-
disintegration
was studied, and the calculation of Ref. \cite{KuAr97}, which treated
PV QE $e$-$d$ scattering. In the latter analysis, only parity-mixing
in the deuteron wave function was considered. In the present study,
we also include the contributions from parity mixing in the final
$e$-$d$ scattering states as well as from PV two-body currents.
Our results are consistent with both of these earlier calculations,
but give a more complete treatment of the QE case.

The remainder of the paper is organized as follows. In Section 
\ref{sec:formalism},
we review the formalism for PV QE scattering and hadronic PV, identify
the relevant operators and matrix elements to be computed, and present
the scaling arguments for the relative magnitude for the hadronic
PV contribution. Section \ref{sec:wavefunctions} gives a discussion
of the bound and scattering state wave functions, which we determine
first in the plane wave approximation and subsequently using the Argonne
$V_{18}$ potential. We present the results of our calculation in
Section \ref{sec:results}, where we consider two cases: threshold
electro-disintegration and QE scattering. Figs. \ref{fig:QEpeak},
\ref{fig:breakdown QE}, \ref{fig:THD}, and \ref{fig:breakdown THD},
which show various contributions to the PV asymmetries as a function
of $Q^{2}$, summarize the main results of this work. A summary discussion
appears in Section \ref{sec:summary}.

\section{PV Electron Scattering and Hadronic PV \label{sec:formalism}}

\subsection{Basic Formalism}

The PV asymmetry for inclusive $e$-$d$ scattering of an unpolarized
target can be expressed in terms of two response functions: $\WEM $,
the parity-conserving (PC) electromagnetic (EM) response, and $\WPV $,
the PV response arising from the interference of EM and PV neutral
current amplitudes. One may decompose the former in terms of the longitudinal
and transverse response functions

\begin{eqnarray}
\WEM  & = & \sum _{f}\left[v_{L}F_{L}^{2}(q)+v_{T}F_{T}^{2}(q)\right]\, 
\Big |_{\omega =E_{f}-E_{i}}\, ,\label{eq:WEM}\\
F_{L}^{2}(q) & = & \sum _{J\geq 0}F_{CJ}^{2}(q)\, ,\\
F_{T}^{2}(q) & = & \sum _{J\geq 1}\left[F_{EJ}^{2}(q)+F_{MJ}^{2}(q)\right]\, ,
\end{eqnarray}
where $v_{L,T}$ are the standard kinematic coefficients (will be
defined later), $q^{\mu }\equiv (\omega ,\vec{q})$ is the four momentum
transfered into the nuclear system ($E_{i}$ and $E_{f}$ are its
initial and final energy). The $F_{XJ}(q)$, $X=C,E,M$, are the charge,
transverse electric, and transverse magnetic multipole matrix elements
depending on $q=\vert \vec{q}\vert $. They are defined through multipole
operators, $\hat{O}^{C}=\hat{M}$, $\hat{O}^{E}=\hat{T}^{el}$, and
$\hat{O}^{M}=i\hat{T}^{mag}$ \cite{dFWa66,Wale75,Mu..94}, as%
\footnote{The extra {}``$i$'' for $\hat{T}^{mag}$ is introduced so that
when real wave functions are used, the form factor $F_{M}$ is real.%
}\begin{equation}
F_{XJ}(q)=\frac{1}{\sqrt{2J_{i}+1}}\sum _{T=0,1}(-1)^{T_{f}-M_{T}}\left(
\begin{array}{ccc}
 T_{f} & T & T_{i}\\
 -M_{T_{f}} & 0 & M_{T_{i}}\end{array}
\right)\langle J_{f},T_{f}\vdots \vdots \hat{O}_{J,T}^{X}(q)\vdots \vdots 
J_{i},T_{i}\rangle \, ,\label{eq:FX}\end{equation}
 where the $\vdots \vdots $ denotes reduced matrix elements in both
angular momentum and isospin \cite{Mu..94}. In the spherical basis,
while a collective quantum label $a$ refers to $(E_{a},L_{a},S_{a},J_{a},
M_{J_{a}},T_{a},M_{T_{a}})$,
the sum $\sum _{f}$ runs over all indexes except $E_{f}$ and $M_{J_{f}}$
because they have been carried out to get Eq. (\ref{eq:WEM}). 

For the PV response, one has \begin{eqnarray}
\WPVZ  & = & \sum _{f}\left[v_{L}W_{AV}^{L}(q)+v_{T}W_{AV}^{T}(q)+v_{T'}
W_{VA}^{T'}(q)\right]\, \Big |_{\omega =E_{f}-E_{i}}\, ,\\
W_{AV}^{L}(q) & = & -\gAe \sum _{J\geq 0}F_{CJ}(q)\tilde{F}_{CJ}(q)\, ,\\
W_{AV}^{T}(q) & = & -\gAe \sum _{J\geq 1}\left[F_{EJ}(q)\tilde{F}_{EJ}(q)+
F_{MJ}(q)\tilde{F}_{MJ}(q)\right]\, ,\\
W_{VA}^{T'}(q) & = & -\gVe \sum _{J\geq 1}\left[F_{EJ}(q)\tilde{F}_{MJ_{5}}
(q)+F_{MJ}(q)\tilde{F}_{EJ_{5}}(q)\right]\, ,
\end{eqnarray}
 where the $\tilde{F}_{X_{(5)}}$ refer to weak, neutral current multipole
matrix elements and the {}``5'' subscript indicates multipole projections
of the axial vector current. The $\tilde{F}_{X_{(5)}}$ are defined
in a similar fashion as Eq. (\ref{eq:FX}) -- up to different coupling
constants, however, for the axial form factors, it is $\hat{M}^{_{5}}$
and $\hat{T}^{el_{5}}$ which have additional factors of $i$ while
$\hat{T}^{mag_{5}}$ is without one \cite{Mu..94}%
\footnote{In Ref. \cite{Mu..94}, $\tilde{F}_{XJ}$ and $\tilde{F}_{XJ_{5}}$
are defined with an extra $1/2$ and $-1/2$ factor. It is found that
the minus sign in Ref. \cite{Mu..94} is a typographical error. The
$1/2$ factor is absorbed here in the overall scale $\ALR ^{0}$. %
}. The kinematic coefficients, $v_{L}$, $v_{T}$, and $v_{T'}$ are
\begin{eqnarray}
v_{L} & = & (Q^{2}/q^{2})^{2}\, ,\\
v_{T} & = & (Q^{2}/q^{2})^{2}/2+\tan ^{2}(\theta /2)\, ,\\
v_{T'} & = & \sqrt{(Q^{2}/q^{2})^{2}+\tan ^{2}(\theta /2)}\, \tan (\theta /2)
\, ,
\end{eqnarray}
where $Q^{2}=q^{2}-\omega ^{2}$, $\theta $ is the scattering angle
of electron.

The PV response functions $W_{AV}^{T,L}$ arise from electron axial
vector ($A$) $\times $ hadronic vector current ($V$) interactions,
while $W_{VA}^{T'}$ is generated by the $V(e)\times A({\textrm{had}.})$
interaction. At tree-level in the Standard Model (SM), the electron
vector coupling to the $Z^{0}$ is suppressed: $\gVe =-1+4\sstw \approx -0.1$
(the axial vector coupling is $\gAe =1$). 

In terms of these response functions, the PV QE asymmetry due to $Z^{0}$-
exchanges
is \begin{equation}
\ALRZ =\frac{G_{\mu }Q^{2}}{4\sqrt{2}\pi \alpha }\frac{\WPVZ }{\WEM }\, .
\label{eq:ALRZ}\end{equation}
 For quasielastic kinematics, $\omega $ and $q$ are related, \emph{viz},
$\omega \approx q^{2}/2\mN $.

\subsection{Hadronic PV Effects}

Hadronic PV effects in the target generate O($\alpha $) corrections
to the tree-level contributions for $W_{VA}^{T'}$. These hadronic
PV effects arise when a photon, rather than a $Z^{0}$, is exchanged
between the electron and hadron. Because the vector $\gamma ee$
coupling $Q^{e}=-1$ is an order of magnitude larger than $\gVe $,
one expects the relative importance of the hadronic PV effects --
compared to the tree-level amplitude -- to be of order 
\begin{equation}
R_{A}\sim-\frac{8\sqrt{2}\pi \alpha }{G_{\mu }\Lambda _{\chi }^{2}}\frac{1}
{1-4\sstw }\frac{g_{\pi }}{g_{A}}\approx -0.01\, ,
\end{equation}
 where $g_{\pi }=3.8\times 10^{-8}$ sets the scale for hadronic PV
interactions, $\Lambda _{\chi }=4\pi F_{\pi }\approx 1$ GeV gives
the scale of chiral symmetry breaking, and $g_{A}=1.267\pm 0.004$.
In the case of elastic $e$-$N$ scattering, hadronic PV arises via
diagrams of the type in Fig. \ref{fig:hadronic PV in eN}.%
\begin{figure}
\includegraphics[  scale=0.7]{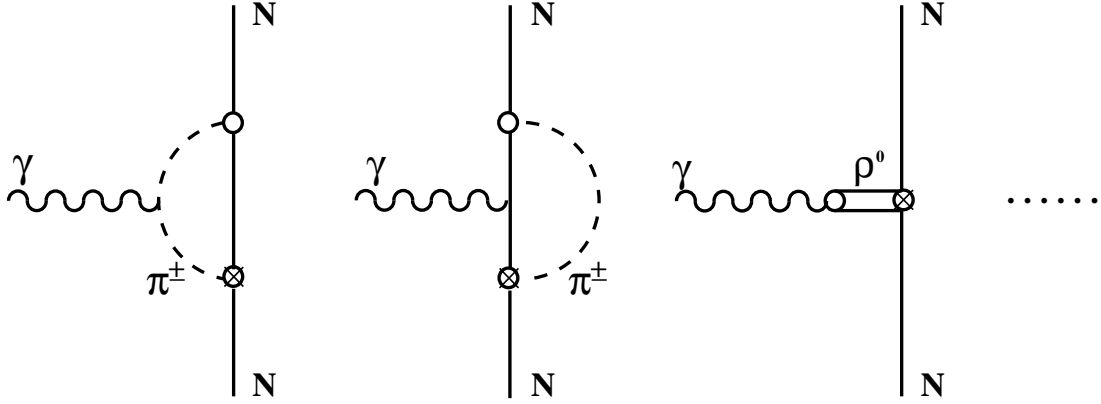}

\caption{Contributions due to hadronic PV in elastic $e$-$N$ scattering,
where $\otimes $ denotes the PV meson-nucleon coupling. \label{fig:hadronic 
PV in eN}}
\end{figure}
 These corrections induce a PV $\gamma NN$ coupling, or anapole moment.
The latter has been used in the one-body estimate of $R_{A}$ given
in Refs. \cite{Zh..00,MuHo90}. Those analyses indicate hadronic PV-induced
anapole corrections of $-6\pm 20\%$.

\begin{figure}
\includegraphics[  scale=0.7]{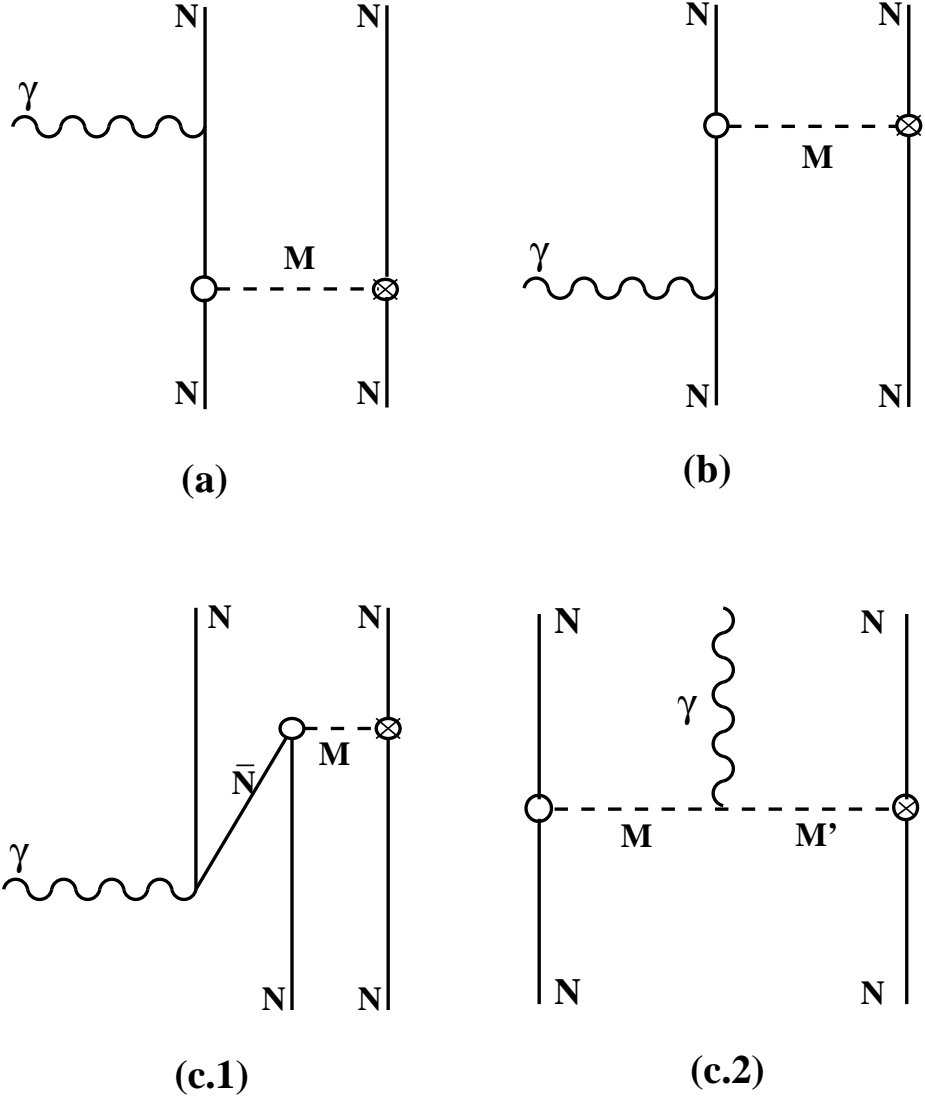}

\caption{Contributions due to two-body hadronic PV in $e$-$d$ scattering.
Here, M and M' denote the identities of mesons. \label{fig:hadronic PV in eD}}
\end{figure}

Two-body hadronic PV contributions arise from the diagrams in Fig.
\ref{fig:hadronic PV in eD}. Figs. \ref{fig:hadronic PV in eD}(a)
and (b) indicate parity-mixing in the initial and final state wave
functions, while Fig. \ref{fig:hadronic PV in eD}(c) indicates the
PV two-body EM current contribution. Each contributes to an effective
axial vector EM transition amplitude, whose effects appear as corrections
to the $\tilde{F}_{XJ_{5}}(q)$ multipole matrix elements. 

In computing the parity-mixing matrix elements, we use the model PV
Hamiltonian given in Ref. \cite{DDH80}: 

\begin{eqnarray}
\HPV (\vec{r}) & = & \frac{g_{\pi NN}h_{\pi }^{1}}{4\sqrt{2}\mN }(i\vect _{1}
\times \vect _{2})_{z}(\vecs _{1}+\vecs _{2})\cdot \vec{u}_{\pi }(\vec{r})-
\nonumber \\
 & - & \frac{g_{\rho NN}}{2\mN }\left(h_{\rho }^{0}\vect _{1}\cdot \vect 2_{2}
+\frac{h_{\rho }^{1}}{2}(\vect _{1}+\vect _{2})_{z}+\frac{h_{\rho }^{2}}
{2\sqrt{6}}(3\tau _{1z}\tau _{2z}-\vect _{1}\cdot \vect _{2})\right)\nonumber 
\\
 & \times  & [(1+\mu _{v})i\vecs _{1}\times \vecs _{2}\cdot \vec{u}_{\rho }
(\vec{r})+(\vecs _{1}-\vecs _{2})\cdot \vec{v}_{\rho }(\vec{r})]\nonumber \\
 & - & \frac{g_{\omega NN}}{2\mN }\left(h_{\omega }^{0}+\frac{h_{\omega }^{1}}
{2}(\vect _{1}+\vect _{2})_{z}\right)\nonumber \\
 & \times  & [(1+\mu _{s})i\vecs _{1}\times \vecs _{2}]\cdot \vec{u}_{\omega }
(\vec{r})+(\vecs _{1}-\vecs _{2})\cdot \vec{v}_{\omega }(\vec{r})]+\nonumber \\
 & - & \frac{1}{4\mN }(\vect _{1}-\vect _{2})_{z}(\vecs _{1}+\vecs _{2})\cdot 
(g_{\omega NN}h_{\omega }^{1}\vec{v}_{\omega }(\vec{r})-g_{\rho NN}h_{\rho }^
{1}\vec{v}_{\rho }(\vec{r}))\nonumber \\
 & - & \frac{g_{\rho NN}}{4\mN }h_{\rho }^{1'}(i\vect _{1}\times \vect _{2})_
{z}(\vecs _{1}+\vecs _{2})\cdot \vec{u}_{\rho }(\vec{r})\, ,\label{eq:HPV}
\end{eqnarray}
 where the $h_{M}^{(X)}$ and $g_{MNN}$ are the weak, PV and strong,
PC meson-nucleon couplings, respectively; and $\vec{u}_{\alpha }(\vec{r})=[
\vec{p}_{1}-\vec{p}_{2},e^{-m_{\alpha }r}/r]$,
$\vec{v}_{\alpha }(\vec{r})=\{\vec{p}_{1}-\vec{p}_{2},e^{-m_{\alpha }r}/r\}$.
Values for the $h_{M}^{(X)}$ have been predicted theoretically using
a variety of approaches. For purposes of our calculation, we will
adopt the so-called DDH {}``best values'' and {}``reasonable ranges\char`\"{}
of Ref. \cite{DDH80}. The latter are consistent with constraints
obtained from a variety of hadronic and nuclear PV experiments, as
discussed in Refs. \cite{AdHa85,HaHo95}.

Current conservation requires that one include contributions from
PV two-body currents to the PV multipole matrix elements. These currents
have been derived in Ref. \cite{HLRM01,HLRM02} from the diagrams
in Fig. \ref{fig:hadronic PV in eD}(c). Complete expressions for
the coordinate space current operators could be found in Ref. \cite{HLRM02}.
To illustrate the structure of these operators, however, we give the
complete two-body current operator associated with the $\pi ^{\pm }$-exchange
component of $\HPV $: 
\begin{eqnarray}
J_{\mu }(\vec{y},\vec{x}_{1},\vec{x}_{2})^{\ssst {PV},\pi } & = & \frac{-eg_{
\pi NN}h_{\pi }^{1}}{8\sqrt{2}\pi \mN }(\vect _{1}\cdot \vect _{2}-\tau _{1z}
\tau _{2z})\bigg (\vecs _{1}\delta ^{(3)}(\vec{y}-\vec{x}_{1})+\vecs _{2}
\delta ^{(3)}(\vec{y}-\vec{x}_{2})\nonumber \\
 &  & -\frac{1}{2}(\vecs _{1}\cdot \vec{\nabla }_{1}-\vecs _{2}\cdot \vec{
\nabla }_{2})(\vec{x}+\frac{1}{2}\frac{x}{m_{\pi }}\vec{\nabla _{y}})(\delta 
^{(3)}(\vec{y}-\vec{x}_{1})+\delta ^{(3)}(\vec{y}-\vec{x}_{2}))\bigg )
\nonumber \\
 &  & \times \frac{e^{-m_{\pi }x}}{x}\quad \mu =1,2,3\nonumber \\
 & \approx  & O(v/c)^2 \quad \mu =0\, ,\label{eq:piMEC}
\end{eqnarray}
where $\vec{x}_{i}$ denotes the position of the $i$-th nucleon and
$x=|\vec{x}|=|\vec{x}_{1}-\vec{x}_{2}|$. Inclusion of these PV two-body
currents guarantees that the the continuity equation, $\vec{\nabla }\cdot 
\vec{J}=i[\hat{H},\rho ]$,
is satisfied at the operator level of the PV NN interaction.

Current conservation may also be implemented in writing down the various
multipole operators. Since nuclear model calculations based on realistic
potentials generally break current conservation, it is useful to implement
the latter via the form of the multipole operator. A well-known example
is Siegert's theorem \cite{Sieg37}, which allows one to rewrite the
$J=1$ transverse electric multipole operator, $\hat{T}_{J=1}^{el}$,
in terms of the electric dipole operator (the $J=1$ charge multiple)
in the long wavelength limit. An extended version of this theorem
\cite{FrFa84} allows one to implement the constraints of current
conservation for transverse electric operators of arbitrary $J$ and
momentum transfer. In general, one has \cite{Musolf89}\begin{equation}
\hat{T}_{J}^{el}(q)=\hat{S}_{J}^{el}(q)+\hat{R}_{J}^{el}(q)\, .\end{equation}
 For $J=1$, one has \begin{eqnarray}
\hat{S}_{1}^{el}(q) & = & \frac{\sqrt{2}}{3}\int d^{3}x\, [\rho (\vec{x})\, 
\vec{x},\hat{H}]\, ,\label{eq:selectric}\\
\hat{R}_{1}^{el}(q) & = & -\frac{q^{2}}{9}\int d^{3}x\, \vec{Y}_{111}^{M_{J}}(
\Omega _{x})\cdot \vec{x}\times \vec{j}.(\vec{x})\, ,\label{eq:relectric}
\end{eqnarray}
where $\vec{Y}_{JL1}^{M_{J}}$ is the vector spherical harmonic.

Note that in the long wavelength limit, $\hat{S}_{1}^{el}(q)$ gives
the leading contribution to $F_{E1}$, $\tilde{F}_{E1}$, and 
$\tilde{F}_{E1_{5}}$.
For elastic electron scattering, hermiticity and time-reversal invariance
require that $F_{E1}$ and $\tilde{F}_{E1}$ must vanish. Moreover,
contributions from $\hat{S}_{1}^{el_{5}}(q)$ also vanish for elastic
scattering, since the commutator in Eq. (\ref{eq:selectric}) leads
to a factor of $\omega =E_{f}-E_{i}=0$. A non-vanishing contribution
arises from $\hat{R}_{1}^{el_{5}}(q)$, whose matrix element constitutes
the nuclear anapole moment contribution. The latter is proportional
to $Q^{2}$ which cancels the $1/Q^{2}$ from the photon propagator
to produce a $Q^{2}$-independent scattering amplitude at lowest order.
This contribution is kinematically indistinguishable from the $Z^{0}$-exchange
contribution to $\tilde{F}_{E1_{5}}$ and, thus, represents a simple
multiplicative correction to the tree-level axial vector response.

For the inelastic transition of interest here, matrix elements of
$\hat{S}_{1}^{el_{5}}(q)$ do not vanish, nor do they contain a power
of $Q^{2}$ to cancel the $1/Q^{2}$ from the photon propagator ($q^{2}\approx 
Q^{2}$
at low energies). The scattering amplitude associated with this operator
goes as $1/Q^{2}$ at low-$Q^{2}$. When multiplied by the factor
of $Q^{2}$ in Eq. (\ref{eq:ALRZ}), this term thus generates a $Q^{2}$-
independent,
non-vanishing contribution to $\ALR $ at low-$Q^{2}$, in contrast
to the $Z^{0}$-exchange asymmetry which vanishes at $Q^{2}=0$. For
sufficiently low-$Q^{2}$, the $Q^{2}$-independent hadronic PV EM
contribution will dominate the asymmetry. As we show below, such a
kinematic region may be reached in principle with threshold PV 
electro-disintegration.
For PV QE scattering at the kinematics of Ref. \cite{SAMPLE00b},
however, the effects of hadronic PV appear to be negligible.

At the kinematics of QE electron scattering, a tower of final state
partial waves contribute to the amplitude, and one must include the
effects of $J>1$ multipole matrix elements (both transverse electric
and transverse magnetic). This situation contrasts with threshold
disintegration, where only the lowest $J$ partial waves may be reached.
As we show in Section \ref{sec:results}, the multipole contributions
to the PV QE asymmetry (due to hadronic PV) saturate for $J\sim 7$.
All multipole matrix elements having $J>1$ carry factors of $Q^{2}$,
so that they do not contribute to the $Q^{2}$-independent term in
the PV asymmetry. Nevertheless, the sum of their effects represents
a tiny correction to the $Z^{0}$-exchange asymmetry.

It is useful to illustrate how PV NN effects contribute to the various
multipole matrix elements entering the axial response, $W_{VA}^{T'}$.
Consider, for example, a transition from the deuteron ground state
to the $^{1}S_{0}$ continuum state. In the absence of the tensor
force component of the strong $NN$ interaction, the deuteron ground
state is pure $^{3}S_{1}$. The PV $NN$ interaction will mix $P$-states
into these $S$-waves. In ordinary perturbation theory, one has

\begin{eqnarray}
\ket {^{3}S_{1}} & \rightarrow  & \ket {^{3}S_{1}}+\ket {\widetilde{
^{3}S_{1}}}\, ,\\
\ket {^{1}S_{0}} & \rightarrow  & \ket {^{1}S_{0}}+\ket {\widetilde{
^{1}S_{0}}}\, ,
\end{eqnarray}
 where the parity mixtures (denoted with a tilde {}``\textasciitilde{}'')
are \begin{eqnarray}
\ket {\widetilde{^{3}S_{1}}} & = & \sum _{k=1,3}\ket {^{k}P_{1}}\frac{
\bra {^{k}P_{1}}\HPV \ket {^{1}S_{0}}}{E_{0}-E_{k}}\, ,\\
\ket {\widetilde{^{1}S_{0}}} & = & \ket {^{3}P_{0}}{\frac{\bra {^{3}P_{0}}
\HPV \ket {^{1}S_{0}}}{E_{0}^{\prime }-E_{1}^{\prime }}}\, .
\end{eqnarray}

For this $J_{i}=1$ to $J_{f}=0$ transition, only $J=1$ multipole
operators contribute. For the PC $\gamma $-exchange contribution,
one has only the magnetic dipole transition between the un-mixed $^{3}S_{1}$
and $^{1}S_{0}$ initial and final state components, resulting in
a non-zero $F_{M1}$ form factor. For the PV $Z^{0}$-exchange amplitude,
only the operator $\hat{T}_{1}^{el_{5}}(q)$ contributes, connecting
the un-mixed $^{3}S_{1}$ and $^{1}S_{0}$ components and leading
to a non-zero ${\tilde{F}}_{E1_{5}}$. The PV NN interaction also
contributes to the latter in three ways: (a) a non-vanishing matrix
element of $\hat{T}_{1}^{el}(q)$ between the initial state $\ket {^{3}S_{1}}$
and the final state $\ket {^{3}P_{0}}$ parity admixture, (b) non-vanishing
matrix elements of $\hat{T}_{1}^{el}(q)$ between the $\ket {^{1,3}P_{1}}$ 
mixture in the initial state and the final state $\ket {^{1}S_{0}}$;
(c) matrix elements of the PV two-body current operator 
${\hat{T}}_{1}^{el_{5}}(q)$
between the $\ket {^{3}S_{1}}$ and $\ket {^{1}S_{0}}$ components.
All other contributions are higher-order in the weak interaction and
can be neglected. The analysis is similar when the $D$-state components
of the deuteron and scattering state induced by the tensor force are
included, as is the analysis for transitions to higher partial waves.

Because the hadronic PV contributes to the asymmetry by inducing axial
photonic couplings, to incorporate these in the expression of Eq.
(\ref{eq:ALRZ}), one only has to modify axial form factors $\tilde{F}_{XJ_{5}}$
by\begin{equation}
\tilde{F}_{XJ_{5}}\rightarrow \tilde{F}_{XJ_{5}}+\beta \, \tilde{F}_{XJ_{5}}^
{\sst {(\gamma )}}\, ,\end{equation}
where\begin{equation}
\beta =-\frac{8\sqrt{2}\pi \alpha }{G_{\mu }Q^{2}}\frac{Q^{e}}{\gVe }\, .
\end{equation}
The EM axial form factors may be decomposed as\begin{eqnarray}
\tilde{F}_{XJ_{5}}^{\sst {(\gamma )}} & = & \frac{1}{\sqrt{2J_{i}+1}}\sum _{
T=0,1}(-1)^{T_{f}-M_{T}}\left(\begin{array}{ccc}
 T_{f} & T & T_{i}\\
 -M_{Tf} & 0 & M_{Ti}\end{array}
\right)\nonumber \\
 & \times  & \left\{ \langle J_{f},T_{f}\vdots \vdots \hat{O}_{J,T}^{X}(q)
\vdots \vdots \widetilde{J_{i},T_{i}}\rangle +\langle \widetilde{J_{f},T_{f}}
\vdots \vdots \hat{O}_{J,T}^{X}(q)\vdots \vdots J_{i},T_{i}\rangle +\langle
 J_{f},T_{f}\vdots \vdots \hat{O}_{J,T}^{X_{5}}(q)\vdots \vdots J_{i},T_{i}
\rangle \right\} \, ,\label{eq:PVmatrixelements}
\end{eqnarray}
 where $\ket {\widetilde{J_{i},T_{i}}}$, $\bra {\widetilde{J_{f},T_{f}}}$,
and $\hat{O}_{J,T}^{X_{5}}(q)$ (the two-body PV EM operators) represent
the effects caused by hadronic PV. One may then express the asymmetry
due to hadronic PV (through the radiative corrections) as\begin{equation}
\ALRg =2\, \frac{\WPVg }{\WEM }\, ,\label{eq:ALRg}\end{equation}
where\begin{equation}
\WPVg =v_{T'}\, \sum _{f}\sum _{J\ge 1}\left[F_{EJ}(q)\tilde{F}_{MJ_{5}}^{
\sst {(\gamma )}}(q)+F_{MJ}(q)\tilde{F}_{EJ_{5}}^{\sst {(\gamma )}}(q)\right]
\, \bigg |_{\omega =E_{f}-E_{i}}\, .\end{equation}

A simple scaling argument allows us to estimate the relative impact
of the two-body hadronic PV contribution. For backward-angle scattering
as studied in the SAMPLE experiments, $v_{T'}\approx v_{T}\gg v_{L}$,
so the ratio of asymmetry due to hadronic PV, Eq. (\ref{eq:ALRg}),
and $Z^{0}$, Eq. (\ref{eq:ALRZ}) is
\begin{equation}
\frac{\ALRg }{\ALRZ }\approx \frac{8\sqrt{2}\pi \alpha }{G_{\mu }Q^{2}}
\frac{\WPVg }{\WPVZ }\approx \frac{8\sqrt{2}\pi \alpha }{G_{\mu }Q^{2}}
\frac{\langle \vec{j}_{\ssst {PV}}^{\gamma }\rangle }{\langle \vec{j}^{Z}
\rangle }\, ,
\end{equation}
While $\expec {\vec{j}^{Z}}$ at backward angles is dominated by the magnetic 
NC component and
scales as $\expec {\vecs }$, $\expec {\vec{j}_{
\ssst {PV}}^{\gamma }}$
scales as $\expec {\vecs }$ times an additional factor introduced
by hadronic PV. Using Eq. (\ref{eq:piMEC}) for a guidance, this factor
is roughly 
\begin{equation}
\frac{g_{\pi NN}h_{\pi }^{1}}{8\sqrt{2}\pi }\left
\langle \frac{e^{-m_{\pi }x}}{\mN x}\right\rangle \, .\end{equation}
With $g_{\pi NN}\cong 13.45$, and $h_{\pi }\sim 4.5\times 10^{-7}$,
the scaling rule is
\begin{equation}
\label{eq:scaling}
\frac{\ALRg }{\ALR }\sim \frac{\mN ^{2}}{Q^{2}}\left\langle \frac{e^{-m_{\pi }
x}}{\mN x}\right\rangle \times 10^{-3}\, ,
\end{equation}
where we have also included the NC magnetic form factor ${\tilde G_M}(Q^2)$
in the denominator. For small $Q^2$, one has ${\tilde G_M}\approx 
\mu_{V}=4.70$.
Taking $x\sim 1$ fm, $\expec {e^{-m_{\pi }x}/(\mN x)}\sim 0.1$, therefore
at $Q^{2}\sim 0.1$ (GeV/$c$)$^{2}$, we have $\ALRg /\ALRZ $ in
the order of a few 0.1\%.

\section{Two-body wave functions \label{sec:wavefunctions}}

In order to compute PV matrix elements in Eq. (\ref{eq:PVmatrixelements}),
we need two-body wave functions. The latter are solutions of the 
Schr{\"o}dinger
equation

\begin{equation}
(H_{\ssst {0}}+\HPV )\ket {\psi +\tilde{\psi }}=E\ket {\psi +\tilde{\psi }}
\, .\label{eq:Schrodingereqn}\end{equation}
Since $\HPV $ is much smaller than $H_{\ssst {0}}$, first-order
perturbation should work well in this process. That is, 
Eq.(\ref{eq:Schrodingereqn})
can be solved in two steps: first, the PC part $\ket {\psi }$ is
determined by solving\begin{equation}
(E-H_{\ssst {0}})\ket \psi =0\, ,\end{equation}
and second, the PV admixture $\ket {\tilde{\psi }}$ is determined
from\begin{equation}
(E-H_{\ssst {0}})\ket {\tilde{\psi }}=\HPV \ket \psi \, ,\label{eq:inhomo}
\end{equation}
with $\ket \psi $ obtained in the first step. In what follows, we
explore two different approaches, one using the plane wave approximation
-- which ignores the final state strong interaction (FSI) -- and one
using a potential model calculation, which includes the FSI.

\subsection{Plane Wave Approximation}

Although the plane wave approximation is na\"{i}ve and simple, we employ
it as a toy-model calculation to achieve some initial insights. In addition, 
the computation of Ref. \cite{KuAr97} employed a plane wave Green's function
to compute the PV admixture in the deuteron, though un-mixed deuteron 
wavefunction
was obtained using the Bonn potential. By comparing the plane wave computation
with the potential model solution (see below), we hope to obtain a sense of 
the 
errors introduced by the plane wave approximation.

In this approach, all the radial components of scattering partial
waves are spherical Bessel function, $j_{L}(pr)$, where $L$ is the
relative orbital angular momentum and $p$ is the relative momentum.

The parity admixture, in first-order perturbation expansion, is expressed
as\begin{equation}
\ket {\tilde{\psi }}=\sum _{\phi }\ket {\phi }\frac{1}{E_{\psi }-E_{\phi }}
\bra {\phi }\HPV \ket {\psi }\, ,\end{equation}
where $\ket {\phi }$ forms a complete eigenbasis. This could be computed
if one knows the Green's function\begin{equation}
G(\vec{x},\vec{y})=\sum _{\phi }\frac{\bra {\vec{x}}\phi \rangle \langle \phi 
\ket {\vec{y}}}{E_{\psi }-E_{\phi }}\, .\end{equation}
 In the plane wave basis, closed-form Green's functions exist for
both calculations of deuteron and final state mixing.

For the deuteron mixing, the transition involves a bound state (binding
energy $E_{B}<0$) to continuum state transition, therefore, 
$E_{B}-E_{\phi }<0$.
The Green's function is\begin{eqnarray}
G^{(\mathcal{D})}(\vec{x},\vec{y}) & = & \sum _{L,S,J,M_{J},T,M_{T}}
\delta _{LST}\left(-\frac{2}{\pi }\gamma \mN \right)\nonumber \\
 & \times  & i_{L}(\gamma r_{<})k_{L}(\gamma r_{>})\mathcal{Y}_{JLS}^{
\dagger M_{J}}(\Omega _{x})\mathcal{Y}_{JLS}^{M_{J}}(\Omega _{y})\chi _{T}^{
\dagger M_{T}}\chi _{T}^{M_{T}}\: ,
\end{eqnarray}
where $\gamma =\sqrt{\mN \vert E_{B}\vert }$; $i_{L}(\gamma r)$
and $k_{L}(\gamma r)$ are the modified spherical Bessel functions
of the first and third kind; $r_{<}\, (r_{>})$ refers to the smaller
(larger) radial coordinate of $x$ and $y$; $\mathcal{Y}$ and $\chi $
denote the spin-angular and isospin wave functions. The factor $\delta _{LST}$,
which is 1 if $L+S+T$ is an odd number and 0 otherwise, enforces
the generalized Pauli principle.

For the final state mixing, because of the pole at $E_{\psi }=E_{\phi }$,
we must add a small imaginary number $\pm i\epsilon $ to the energy
denominator, as in the scattering problem. In this way, we obtain
a retarded (advanced) Green's function corresponding to the $-i\epsilon $
($+i\epsilon $) prescription. However, only the real part of this
Green's function gives a non-vanishing response function. The real
part is equivalent to the average of retarded and advanced ones 
\begin{eqnarray}
\bar{G}^{(\mathcal{F})}(\vec{x},\vec{y}) & = & \sum _{L,S,J,M_{J},T,M_{T}}
\delta _{LST}(\mN p_{\ssst {\mathcal{F}}})\nonumber \\
 & \times  & j_{L}(p_{\ssst {\mathcal{F}}}r_{<})n_{L}(p_{\ssst {\mathcal{F}}
}r_{>})\mathcal{Y}_{JLS}^{\dagger M_{J}}(\Omega _{x})\mathcal{Y}_{JLS}^{M_{J}}
(\Omega _{y})\chi _{T}^{\dagger M_{T}}\chi _{T}^{M_{T}}\: ,
\end{eqnarray}
 where $p_{\ssst {\mathcal{F}}}$ is the relative momentum of the
final state and $n_{L}(p_{\ssst {\mathcal{F}}}r)$ is the spherical
Neumann wave functions.

Illustrative results for the plane wave calculation are given in Figs. 
\ref{fig:PC comparison} and \ref{fig:D PNC}. 
We note that, in comparison with the complete, coupled 
channel potential model computation 
(see below), use of the plane wave Green's function overestimates the 
degree of parity-mixing in the deuteron ground state. For parity-mixing
in the scattering states, we also find a mismatch between the two 
approaches, though no systematic pattern emerges as to the magnitude or
sign of the difference. The problem may be particularly severe for the 
$^3$S$_1$ and $^3$D$_1$  scattering states which, in the plane wave 
approach, are not automatically orthogonal to the deuteron wave function
$|{\cal D}\rangle$. Although one might attempt to solve this problem by
implement orthogonality by hand, {\em viz},
\begin{equation}
\ket {^{3}S_{1},^{3}D_{1}}_{\perp }=\ket {^{3}S_{1},^{3}D_{1}}-\ket {
\mathcal{D}}\langle \mathcal{D}\ket {^{3}S_{1},^{3}D_{1}}\, ,
\end{equation}
it is questionable whether this {\em ad hoc} solution is rigorously 
correct. For these reasons, then, we rely only on the coupled channel
potential model computation to determine the nuclear PV contribution to 
the inelastic asymmetry.

\subsection{Potential Model Calculation}

Although the $NN$ potential determined directly from solving QCD
is not available, a variety of modern phenomenological potentials
successfully fit $NN$ scattering data (below 350 MeV or so) as well
as deuteron properties with reasonable $\chi ^{2}$ values. Here,
we use the Argonne $V_{18}$ potential (AV$_{18}$) \cite{AV18}.

The PC scattering wave function, \begin{equation}
\bra {\vec{r}}E,L,S,J,M_{J},T,M_{T}\rangle =\sqrt{\frac{2\mN k}{\pi }}
\frac{u_{\ssst {JLS}}(r)}{r}\, \mathcal{Y}_{\ssst {JLS}}^{\ssst {M_{J}}}\, 
\chi _{T}^{M_{T}}\, ,\end{equation}
where $u(r)$ denotes the radial wave function; $k=\sqrt{\mN E}$;
and the overall constant is fixed by the normalization condition 
$\langle E'...\vert E...\rangle =\delta (E'-E)...$,
is determined by solving the Schr{\"o}dinger equation. This task
is eventually reduced to integrating a one-dimensional differential
equation for the radial component and solving for the phase shift.
However, due to the tensor force, for $J>0$, states having quantum
numbers $(L,S,J)=(J-1,1,J)$ and $(J+1,1,J)$ are coupled, requiring
that one solve a coupled set of differential equations. The normalization
for radial wave functions are fixed by their asymptotic forms. For
the uncoupled channel problem, one has

\begin{equation}
u_{\ssst {JLS}}(r\rightarrow \infty )=r\, \sin (kr+L\pi /2+\delta _{
\ssst {JLS}})\, ,\end{equation}
where $\delta _{\ssst {JLS}}$ denotes the phase shift. For the coupled
channel problem, the convention introduced by Blatt and Biedenharn
(BB) \cite{BlBi52}, with two eigenphases shifts $\delta _{\ssst {J}}^{(1)}$,
$\delta _{\ssst {J}}^{(2)}$ and a mixing parameter $\epsilon _{\ssst {J}}$,
is adopted.%
\footnote{The {}``nuclear bar'' convention, defined in Ref. \cite{StYM57},
is more commonly used in literature. However, the phase parameters of these
two conventions are totally interchangeable. The reason for our choice
is that all the wave functions are purely real in BB convention.%
} The two orthogonal, real solutions are\begin{eqnarray}
\left(\begin{array}{c}
 u_{\ssst {L=J-1}}^{(1)}\\
 u_{\ssst {L=J+1}}^{(1)}\end{array}
\right)(r\rightarrow \infty ) & = & r\, \left(\begin{array}{c}
 \cos \epsilon _{\ssst {J}}\, \sin (kr+(J-1)\pi /2+\delta _{\ssst {J}}^{(1)})\\
 \sin \epsilon _{\ssst {J}}\, \sin (kr+(J+1)\pi /2+\delta _{\ssst {J}}^{(1)})
\end{array}
\right)\, ,\\
\left(\begin{array}{c}
 u_{\ssst {L=J-1}}^{(2)}\\
 u_{\ssst {L=J+1}}^{(2)}\end{array}
\right)(r\rightarrow \infty ) & = & r\, \left(\begin{array}{c}
 -\sin \epsilon _{\ssst {J}}\, \sin (kr+(J-1)\pi /2+\delta _{\ssst {J}}^{(2)})
\\
 \cos \epsilon _{\ssst {J}}\, \sin (kr+(J+1)\pi /2+\delta _{\ssst {J}}^{(2)})
\end{array}
\right)\, .
\end{eqnarray}
It should be noted that while we will still refer to solution 1(2)
as $^{3}[J-1]_{J}$($^{3}[J+1]_{J}$) state, it contains a component
involving the other channel. We have verified our calculations by
reproducing the experimental phase shifts.

The deuteron wave function,

\begin{equation}
\bra {\vec{r}}\mathcal{D},M_{J}\rangle =\left\{ \frac{u(r)}{r}\, \mathcal{Y}_{
\ssst {101}}^{\ssst {M_{J}}}+\frac{w(r)}{r}\, \mathcal{Y}_{\ssst {121}}^{
\ssst {M_{J}}}\right\} \, \chi _{0}^{0}\, ,\end{equation}
is obtained by solving the eigenvalue problem for binding energy $E_{\sst {B}}$
and $D/S$ ratio. The asymptotic and normalization conditions are
\begin{eqnarray}
 &  & u(r\ll 1)\propto r\, ,\quad u(r\gg 1)\propto r\, k_{\ssst {0}}(\gamma r)
\, \\
 &  & w(r\ll 1)\propto r^{3}\, ,\quad w(r\gg 1)\propto r\, k_{\ssst {2}}(
\gamma r)\, \\
 &  & \int dr\, [u^{2}(r)+w^{2}(r)]=1\, .
\end{eqnarray}

Although one can follow a similar strategy and obtain the PV wave
functions by the Green's function method mentioned in previous subsection,
it is not straightforward to do so; the unperturbed wave functions
are too complex to allow one to obtain analytical results as in the
case of plane waves. Therefore, following the same approach as in
Ref. \cite{HwHM81}, we directly solve the inhomogeneous equation,
Eq. (\ref{eq:inhomo}).

The basic idea is to solve the problem twice, once with the {}``source''
term off (thus a homogeneous equation as solving the PC wave function)
and then with the source term on. A general solution for the inhomogeneous
equation, $\tilde{\psi }_{g}$, can be expressed as a linear combination
of solutions for the homogeneous equation, called the complimentary
solutions, $\tilde{\psi }_{c}(i)$, plus the particular solution,
$\tilde{\psi }$. Therefore, in order to obtain the particular solution,
\begin{equation}
\tilde{\psi }=\tilde{\psi }_{g}-\sum _{i}\alpha _{i}\psi _{c}(i)\, ,
\label{eq:particularsol.}\end{equation}
the complimentary part has to be fully subtracted. Thus, we must determine
$\alpha _{i}$, $i=1$...N, N being the number of coupled equations.

In the case of solving scattering wave function, the asymptotic behaviors
of both $\tilde{\psi }_{g}$ and $\tilde{\psi }_{c}$ can be expressed
as linear combinations of incoming and outgoing spherical waves. While
the interactions cause phase shifts of outgoing waves, the incoming
waves are not altered. This observation tells us that, $\tilde{\psi }$,
the parity-mixed component induced by the PV $NN$ interaction, should
not contain any incoming component. Using this result, $\alpha _{i}$
are solution when the incoming wave components of $\tilde{\psi }_{g}$
and $\tilde{\psi }_{c}(i)$ completely cancel in Eq.(\ref{eq:particularsol.}).

Except for $^{1}S_{0}$ and $^{3}P_{0}$, which can only be mixed
into each other, all the other uncoupled states, $^{1,3}L_{J=L}$
could have mixtures from $^{3}[L-1]_{L}$ and $^{3}[L+1]_{L}$ states.
For the coupled states, $^{3}[L=J-1]_{J}$ and $^{3}[L=J+1]_{J}$,
both mix to $^{1,3}J_{J}$. If the mixture is an uncoupled state,
and we have\begin{eqnarray}
\tilde{\psi }_{c} & \rightarrow  & a\, e^{-i(kr-L\pi )}+b\, e^{i(kr-L\pi )}
\, ,\\
\tilde{\psi }_{g} & \rightarrow  & c\, e^{-i(kr-L\pi )}+d\, e^{i(kr-L\pi )}\, ,
\end{eqnarray}
then\begin{equation}
\tilde{\psi }=\tilde{\psi }_{g}-\frac{c}{a}\tilde{\psi }_{c}\, .\end{equation}
When the mixtures are coupled, a two channel calculation is needed.
If one has

\begin{eqnarray}
\tilde{\psi }_{c}(i=1,2) & \rightarrow  & \left(\begin{array}{c}
 a(i,1)e^{-i(kr-(J-1)\pi )}+b(i,1)e^{i(kr-(J-1)\pi )}\\
 a(i,2)e^{-i(kr-(J+1)\pi )}+b(i,2)e^{i(kr-(J+1)\pi )}\end{array}
\right)\, ,\\
\tilde{\psi }_{g}(i=1,2) & \rightarrow  & \left(\begin{array}{c}
 c(i,1)e^{-i(kr-(J-1)\pi )}+d(i,1)e^{i(kr-(J-1)\pi )}\\
 c(i,2)e^{-i(kr-(J+1)\pi )}+d(i,2)e^{i(kr-(J+1)\pi )}\end{array}
\right)\, ,
\end{eqnarray}
then\begin{eqnarray}
\tilde{\psi } & = & \tilde{\psi }_{g}(1)-\alpha _{1}(1)\tilde{\psi }_{c}(1)-
\alpha _{2}(1)\tilde{\psi }_{c}(2)\, ,\\
 & = & \tilde{\psi }_{g}(2)-\alpha _{1}(2)\tilde{\psi }_{c}(1)-\alpha _{2}(2)
\tilde{\psi }_{c}(2)\, ,
\end{eqnarray}
with\begin{equation}
\left(\begin{array}{c}
 \alpha _{1}(i)\\
 \alpha _{2}(i)\end{array}
\right)=\left(\begin{array}{cc}
 a(1,1) & a(2,1)\\
 a(1,2) & a(2,2)\end{array}
\right)^{-1}\left(\begin{array}{c}
 c(i,1)\\
 c(i,2)\end{array}
\right)\, .\end{equation}
Note that all the coefficients here are complex. This implies that
the mixed wave functions are also complex. However, in our framework,
only the real part will contribute to the response function $\WPVg $
and, thus, to the asymmetry. 

Various criteria exist for testing the numerical solutions: i) they
should satisfy the differential equation, ii) they should be independent
of the initial conditions used to integrate the differential equation,
iii) they should be proportional to the source term, i.e., if the
source term doubles, the solution should also double. These conditions
are employed to make sure we obtain the correct solutions.

As for the parity admixture of deuteron, it is determined using the
same procedure. Since one is dealing with a bound state however, the
asymptotic behavior is given by a linear combination of modified spherical
Bessel functions, $i_{L}$ and $k_{L}$. The physically realistic
solution is obtained by completely subtracting the $i_{L}$ component,
because it exponentially diverges.

\section{Results and discussion \label{sec:results}}

\begin{figure}
\subfigure[$^1S_0$ channel]{\includegraphics{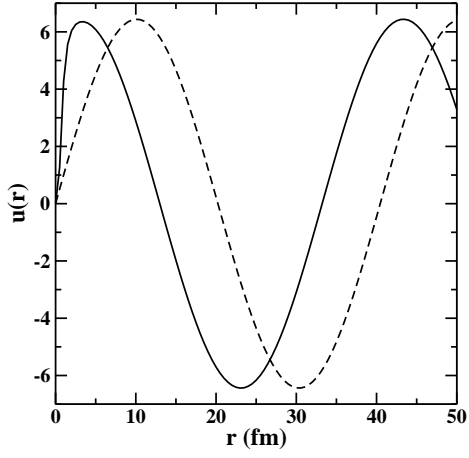}}\subfigure[$^3P_1$ 
channel]{\includegraphics{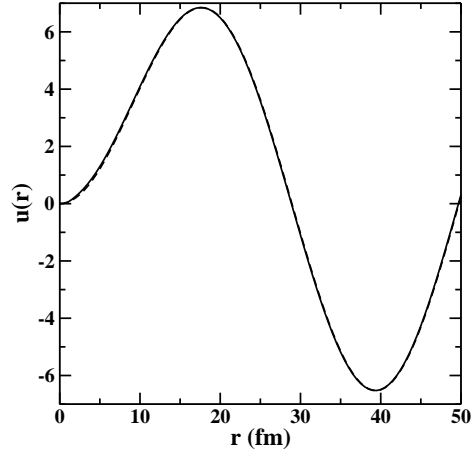}}

\caption{Comparison of scattering state wave functions: dashed lines give
the plane wave solutions and solid lines give results of the potential
model calculations using AV$_{18}$. The relative $np$ energy
is 1 MeV.\label{fig:PC comparison} }
\end{figure}

\begin{figure}
\subfigure[$^3P_0$ admixture in $^1S_0$]{\includegraphics{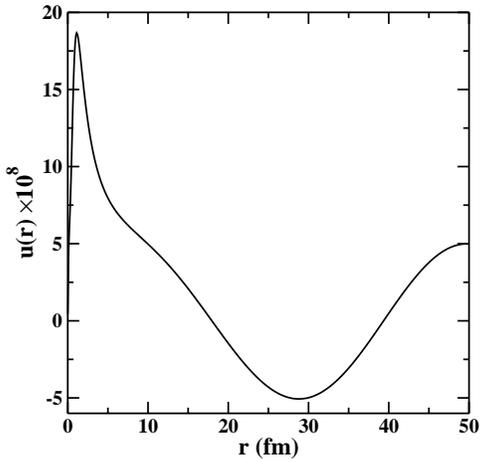}}
\subfigure[$^1S_0$ admixture in $^3P_0$]{\includegraphics{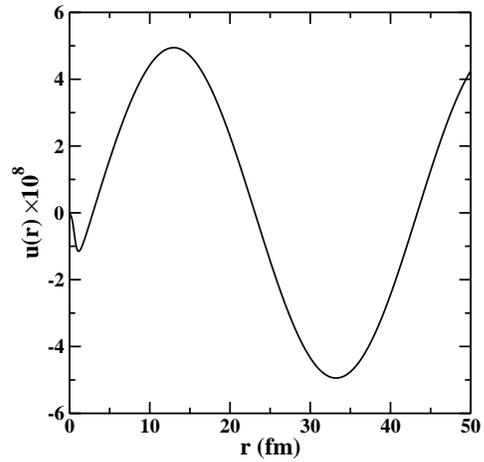}}

\caption{PV admixtures for the scattering states in Fig. 
\ref{fig:PC comparison},
obtained by solving the inhomogeneous differential equations. 
\label{fig:F PNC}}
\end{figure}

\begin{figure}
\subfigure[$^1P_1$ admixture]{\includegraphics{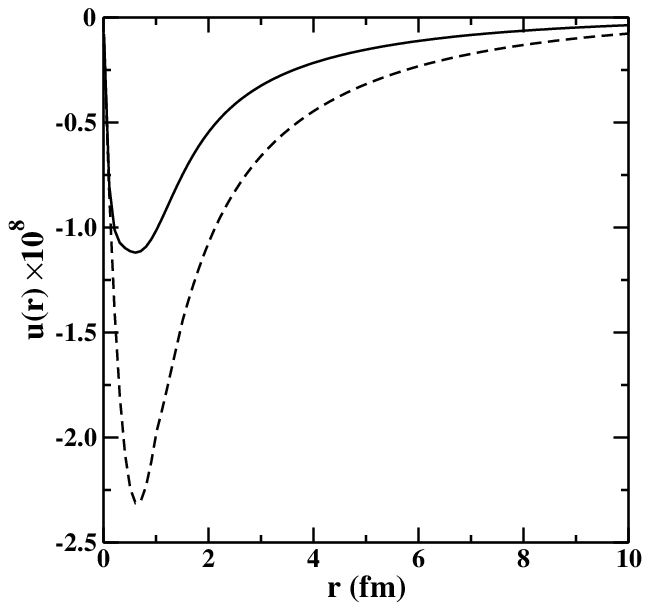}}
\subfigure[$^3P_1$ admixture]{\includegraphics{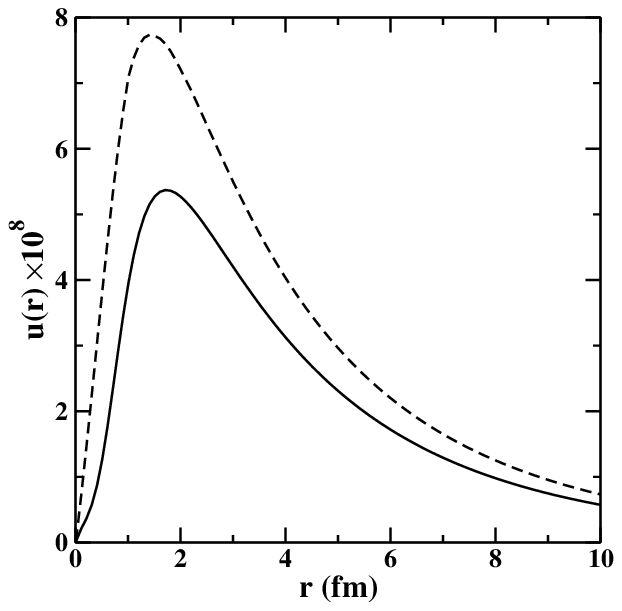}}

\caption{PV admixtures of deuteron: dashed lines are results using the plane
wave Green's function and solid lines are calculations using AV$_{18}$.
\label{fig:D PNC} }
\end{figure}

First, we compare the two approaches discussed in Section 
\ref{sec:wavefunctions}.
Fig. \ref{fig:PC comparison} shows an example of the comparison between
the plane wave scattering states and the ones obtained from the potential
model calculations (in our case, it is AV$_{18}$). Though at $E_{rel}=1$
MeV, the $^{3}P_{0}$ solutions look almost the same, the plane wave
$^{1}S_{0}$ state differs from the more realistic solution by a large
phase shift as well as in its radial shape at small distances. Note
that the latter difference is important because the PV $NN$ interaction
is very sensitive to the short range behavior of wave functions. Therefore,
the plane wave approximation is not adequate. Fig. \ref{fig:F PNC}
shows the PV mixtures for these $^{1}S_{0}$ and $^{3}P_{0}$ states,
$^{3}P_{0}$ and $^{1}S_{0}$ respectively. They are similar to the
results of Ref. \cite{HwHM81}, which were obtained by using Reid
soft-core potential, but slightly differ in magnitudes. For the deuteron
mixing, the $^{1}P_{1}$ state is induced only by $\rho $ and $\omega $
exchanges ($\Delta T=0$), while the $^{3}P_{1}$ is induced dominantly
by the $\pi $ exchange; both results are plotted in Fig. \ref{fig:D PNC}.
Also shown by the dotted lines in the same figure are the solutions
of the plane wave Green's function. Though these curves are similar
in shape, the potential model calculation gives smaller amplitudes
and different small-$r$ radial dependence than plane waves. From
now on, we only present results from the potential model calculation
which is more realistic. 

\begin{figure}
\subfigure[Cross section]{\includegraphics[  scale=0.3,
  angle=270,
  origin=lB]{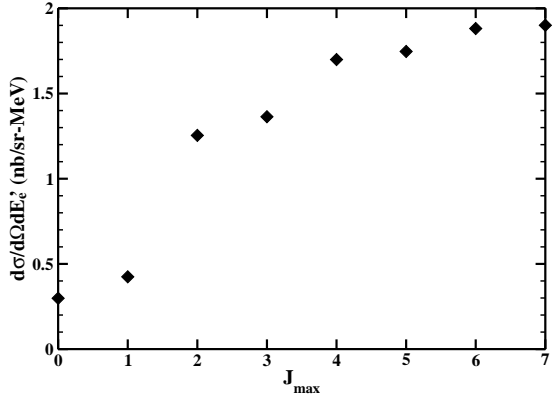}}\subfigure[Asymmetry]{\includegraphics[  scale=0.3,
  angle=270,
  origin=lB]{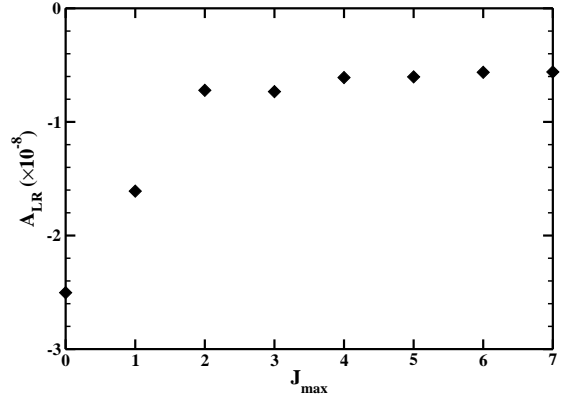}}

\caption{The saturation behavior of total cross section and asymmetry as 
functions
of $J_{max}$, the maximum total angular momentum for final states
being included. \label{fig:saturation}}
\end{figure}

As the impact of $\GAeV $ on $\ALR $ is more important at backward
angles, we first examine the extreme case: $\theta =180^{\circ }$.
Subsequently, we present results relevant to SAMPLE kinematics. The
maximum $Q^{2}$ we consider is 0.15 (GeV/\emph{c)$^{2}$,} and the
saturation behavior shown in Fig. \ref{fig:saturation} justifies 
truncation of the sum over final scattering states at total angular
momenta $J_{f}\le 7$. 

\begin{figure}
\subfigure[Cross section]{\includegraphics[  scale=0.3,
  angle=270,
  origin=lB]{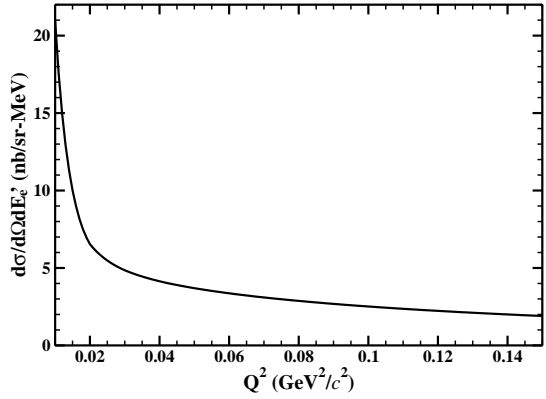}}\subfigure[Asymmetry]{\includegraphics[  
scale=0.3,
  angle=270,
  origin=lB]{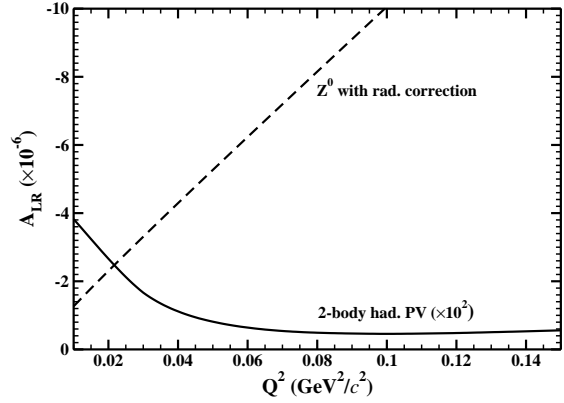}}

\caption{The total cross section and PV asymmetry versus $Q^{2}$, where the
kinematics are constrained to satisfy $q^{2}=2\mN \omega $ and $\theta =180^{
\circ }$.
Note in (b), the hadronic PV contribution is multiplied by 100; it
does not actually cross the $Z^{0}$ term in this $Q^{2}$ range.
\label{fig:QEpeak}}
\end{figure}

Fig. \ref{fig:QEpeak} indicates how the backward angle cross section
and asymmetry vary with $Q^{2}$, ranging from 0.01 to 0.15 (GeV/$c$)$^{2}$
at the QE peak. It is clear that the asymmetry due to two-body hadronic
PV (plotted with a magnification of 100) is insignificant compared
with the contribution of tree-level $Z^{0}$-exchange plus radiative
corrections, which includes nucleon anapole effects. The curve for
$Z^{0}$-exchange asymmetry, plotted using the static approximation
result in Ref. \cite{Mu..94} with parametrized nucleon form factors,
shows the expected proportionality to $Q^{2}$. The curve for hadronic
PV shows a 0.05\% correction to $\ALRZ $ at $Q^{2}=0.1$ and a 0.3\%
correction at $Q^{2}=0.04$. We note that these results are consistent with 
simple scaling arguments as Eq. (\ref{eq:scaling}). Although there is 
some enhancement for
$\ALRg $ as $Q^{2}$ decreases, even at $Q^{2}\sim 0.01$, near the
threshold for QE kinematics, the correction is less than 5\%.

\begin{figure}
\includegraphics[  scale=0.5,
  angle=270,
  origin=lB]{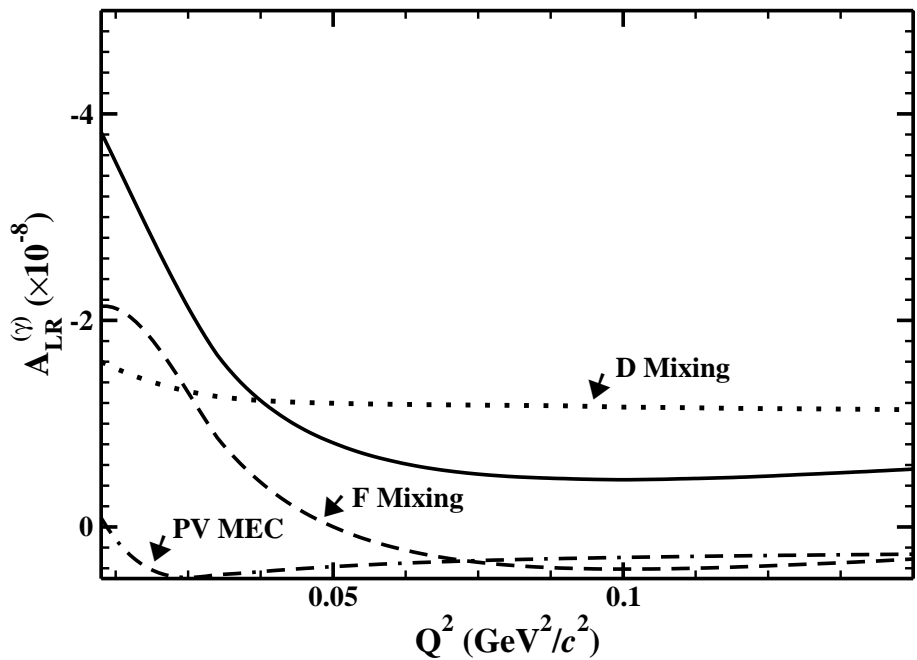}

\caption{The breakdown of various hadronic PV contributions to the asymmetry
at QE kinematics, where D, F, and MEC refer to contributions from
deuteron mixing, final state mixing, and PV meson exchange currents,
respectively, and the solid line gives the total. \label{fig:breakdown QE}}
\end{figure}

A detailed breakdown of various hadronic PV contributions is shown
in Fig. \ref{fig:breakdown QE}. The deuteron mixing, rather insensitive
to the $Q^{2}$ of the explored region, is the dominant contribution
for $Q^{2}\ge 0.03$. Its correction to $\ALRZ $ at $Q^{2}=0.1$
is approximately 0.1\%, and 0.3\% at $Q^{2}=0.04$. These values are
consistent in the order of magnitude with the results of Ref. \cite{KuAr97},
which used Bonn potential to calculate the PC wave functions and the
plane wave Green's function to calculate the parity mixture in deuteron.
On the other hand, the final state mixing and PV meson exchange currents,
though comparatively smaller, do have a combined contribution which
could be as large as half of the contribution from deuteron mixing
for $Q^{2}\ge 0.03$. They are also more sensitive to $Q^{2}$ and
become important when approaching the QE threshold.

\begin{figure}
\subfigure[Cross section]{\includegraphics[  scale=0.3,
  angle=270,
  origin=lB]{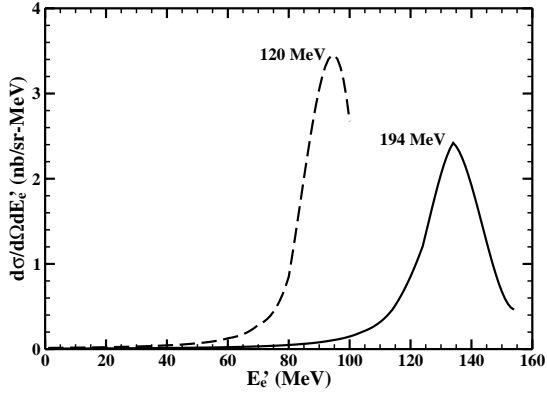}}\subfigure[Asymmetry]{
\includegraphics[  scale=0.3,
  angle=270,
  origin=lB]{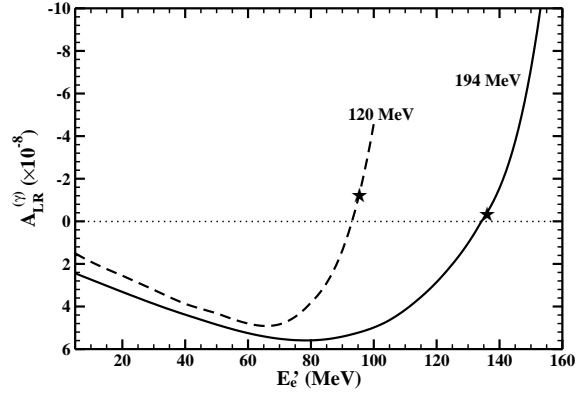}}

\caption{The total cross section and two-body hadronic PV asymmetry versus
electron final energy, for 194 and 120 MeV incident beams and $180^{\circ }$
scattering angle. The asterisk denotes the position of the QE peak.
\label{fig:SAMPLEtheta180}}
\end{figure}
\begin{figure}
\subfigure[194 MeV beam]{\includegraphics[  scale=0.3,
  angle=270,
  origin=lB]{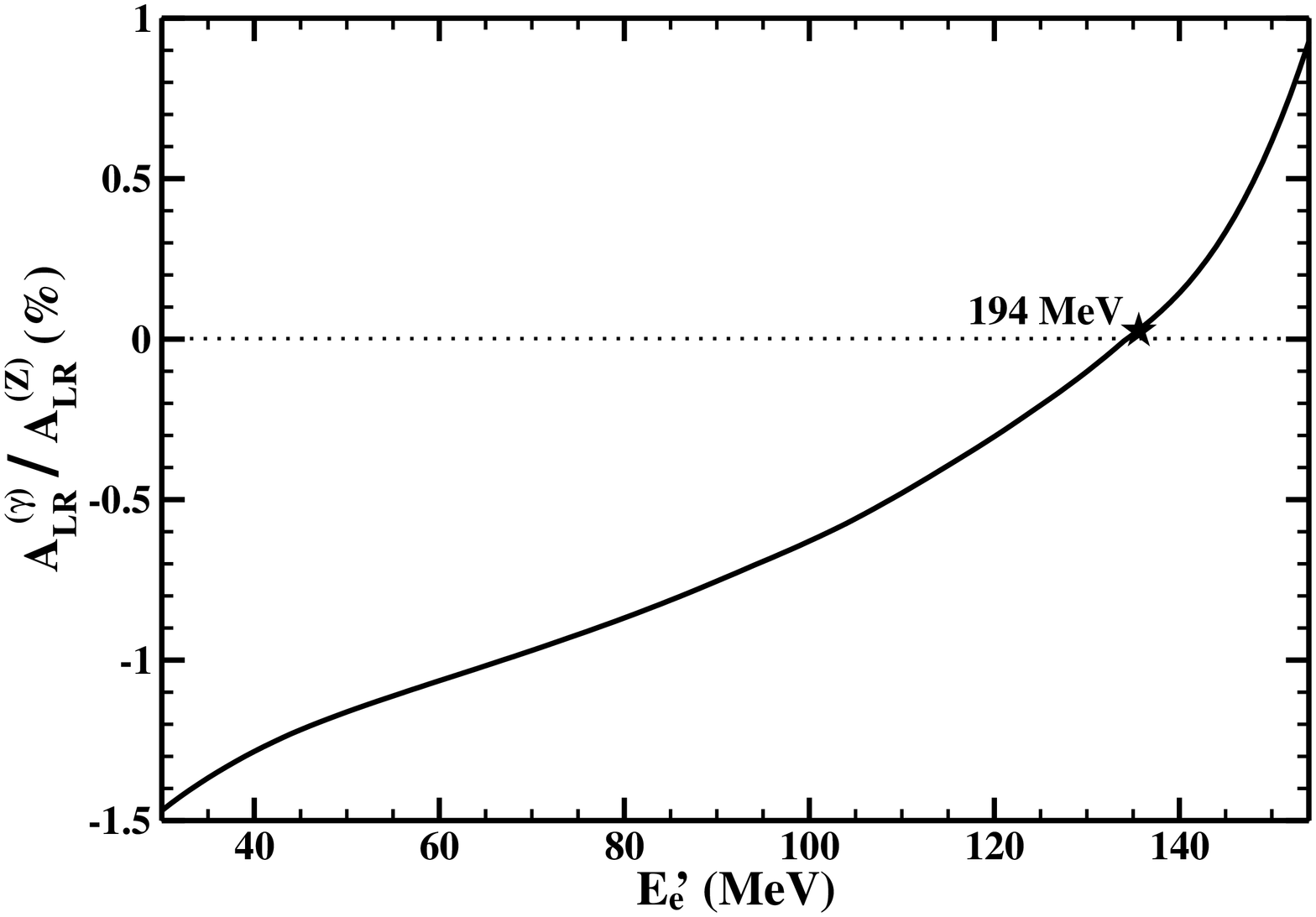}}\subfigure[120 MeV beam]{
\includegraphics[  scale=0.3,
  angle=270,
  origin=lB]{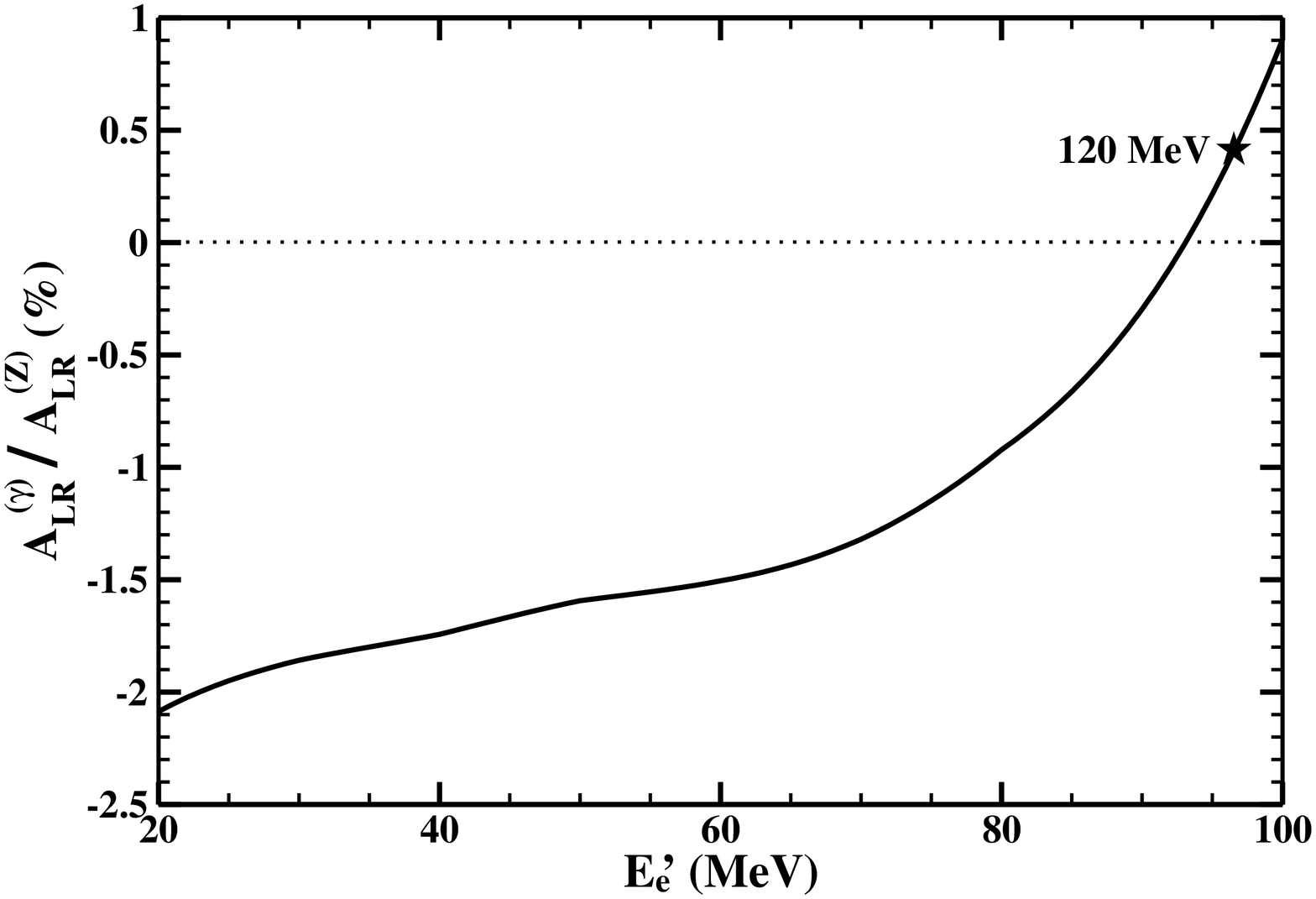}}

\caption{The ratio of asymmetry due to two-body hadronic PV and $Z^{0}$-
exchange,
versus final electron energy. Both kinematics are the same as in Fig.
\ref{fig:SAMPLEtheta180}. \label{fig:SAMPLEtheta180ratio}}
\end{figure}

Away from the QE peak, the dependence of the cross section and asymmetry
on final electron energy are shown in Fig. \ref{fig:SAMPLEtheta180}
for 194 and 120 MeV beams. Since the scattered electrons are detected
via the {\v C}erenkov radiation (the threshold is about 20 MeV), scattered
electrons with $E_{e}'\lesssim $ 150 and 100 MeV, respectively, are
detected. However, judging from the cross section plot, only regions
about peak energy $\pm $ 20 and 10 MeV, respectively, are important
for these two cases. When these asymmetries are further plotted as
ratios to $\ALRZ $, as shown in Fig. \ref{fig:SAMPLEtheta180ratio},
we observe that the correction could become as large as a few percent.
Notice, however, that the corrections change sign roughly when crossing
the QE ridge. Hence, corrections from these two regions cancel after
integration, thereby, keeping the total correction small. A similar
feature was also found in the calculation of Ref. \cite{DSvK00},
where PC two-body effects were considered.

\begin{figure}
\begin{tabular}{cc}
\includegraphics[  scale=0.3,
  angle=270,
  origin=lB]{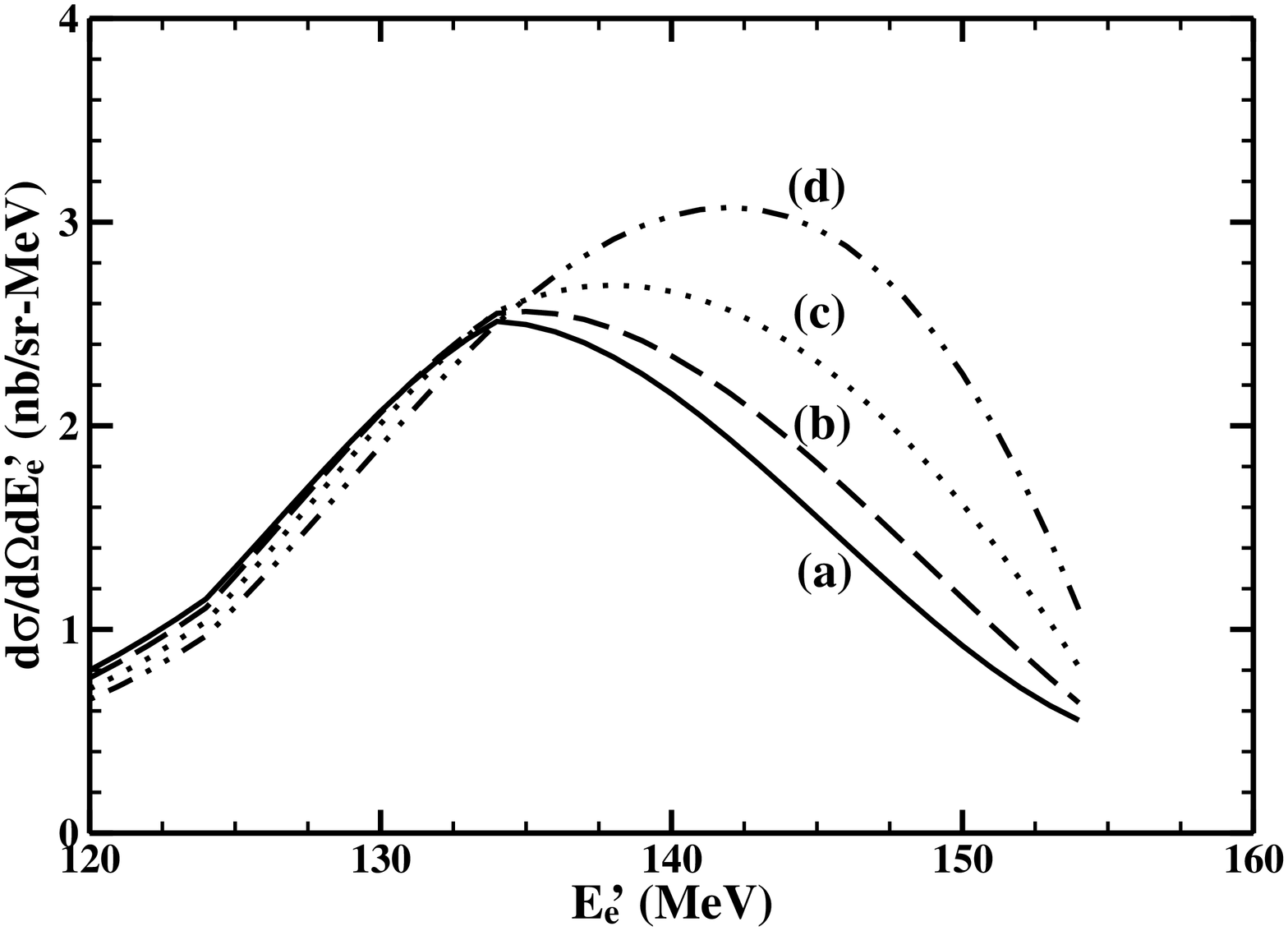}&
\includegraphics[  scale=0.3,
  angle=270,
  origin=lB]{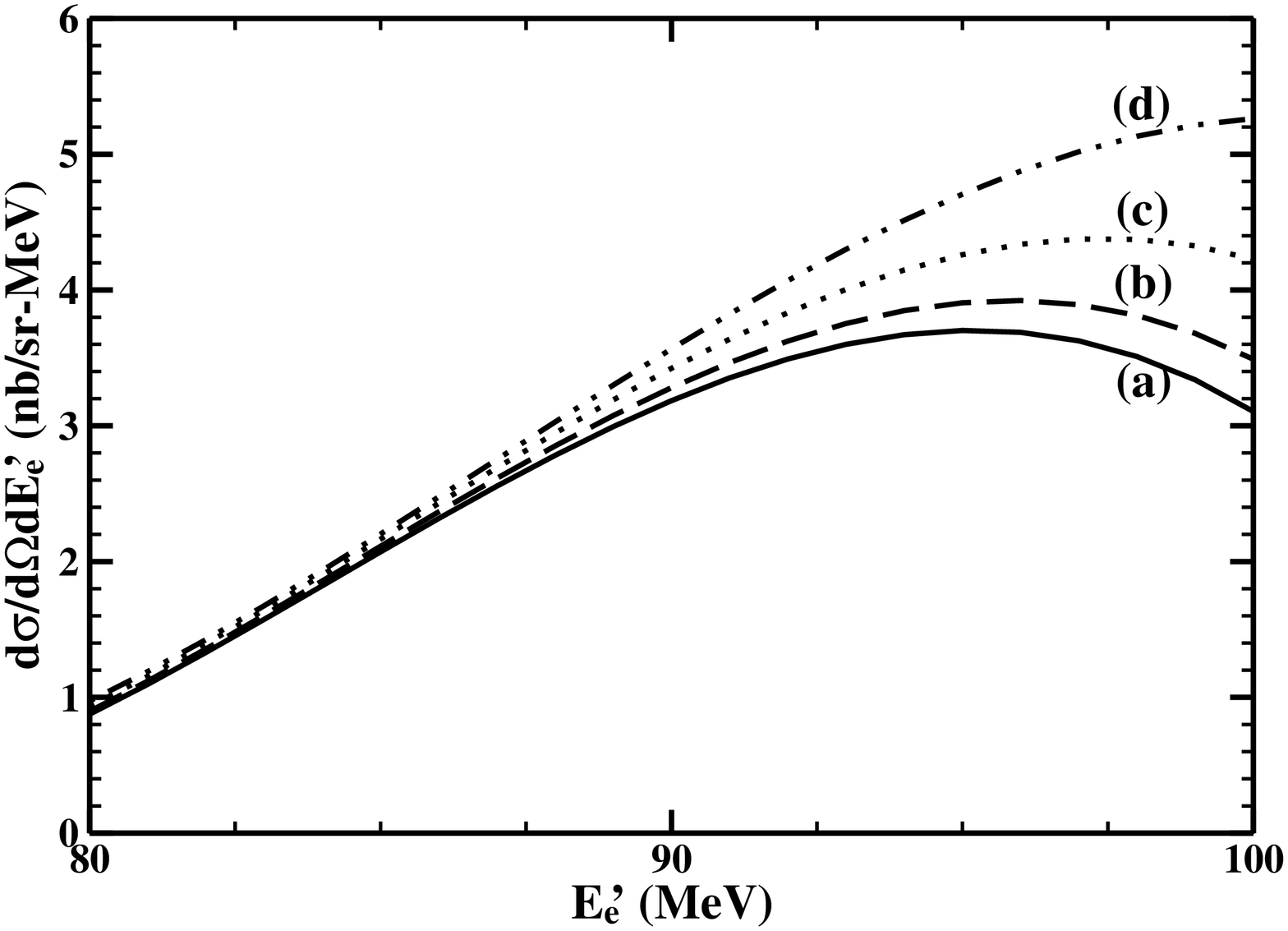}\\
\includegraphics[  scale=0.3,
  angle=270,
  origin=lB]{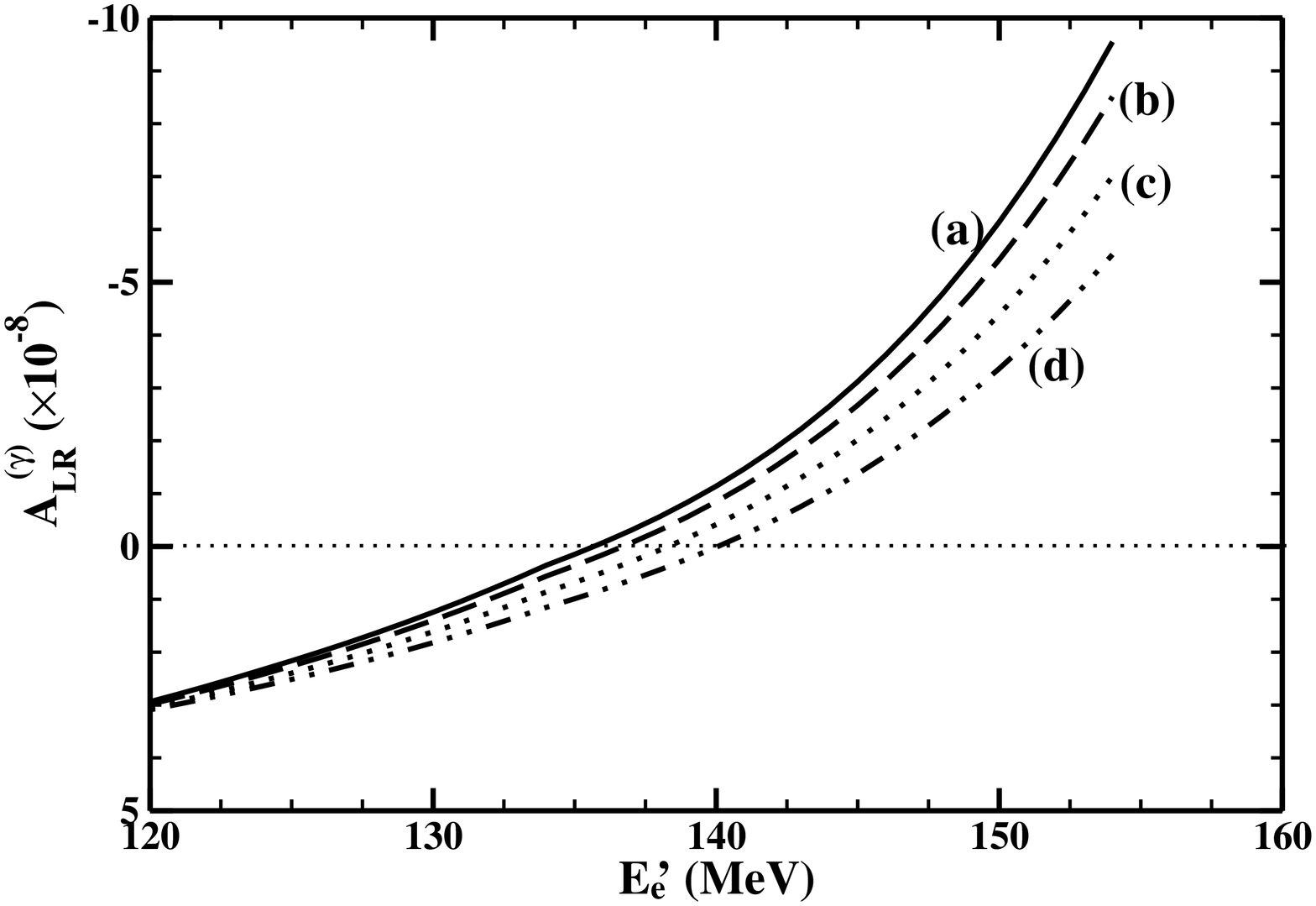}&
\includegraphics[  scale=0.3,
  angle=270,
  origin=lB]{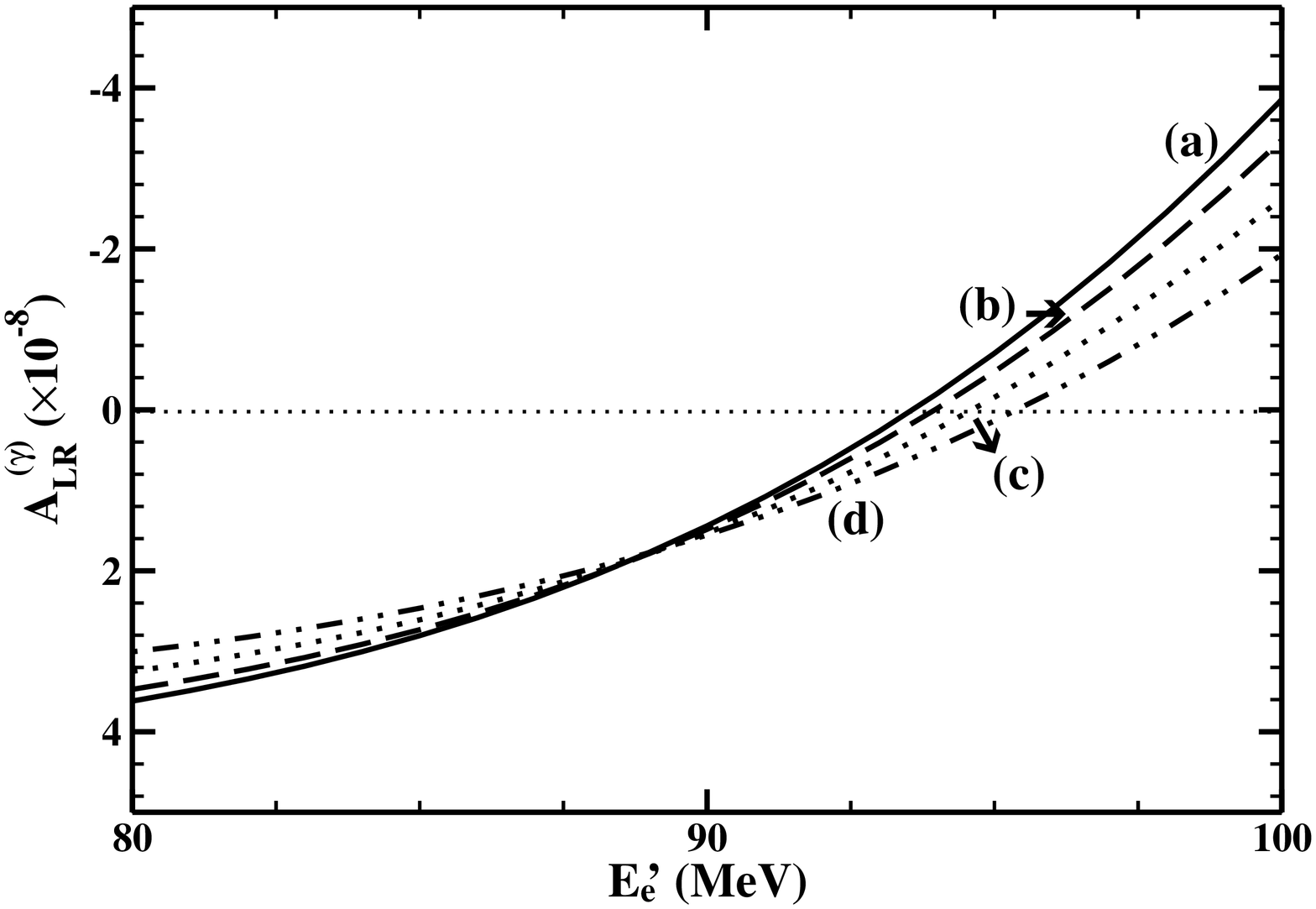}\\
\end{tabular}

\caption{The same plots as Fig. \ref{fig:SAMPLEtheta180} with four average
detector angles of SAMPLE experiments: (a) $160.5^{\circ }$, (b)
$154.0^{\circ }$, (c) $145.9^{\circ }$, and (d) $138.4^{\circ }$.
The left panels are for the $E_{e}=194$ MeV and right panels for
the $E_{e}=120$ MeV case. \label{fig:SAMPLE real}}
\end{figure}

The setup of SAMPLE experiments actually cover the angular range from
$130^{\circ }$ to $170^{\circ }$, the average angles of the detectors
are: $138.4^{\circ }$, $145.9^{\circ }$, $154.0^{\circ }$, and
$160.5^{\circ }$ \cite{TIto01}. The corresponding cross sections
and asymmetries are plotted in Fig. \ref{fig:SAMPLE real}. The general
trend is that when the angle gets smaller, the cross section becomes
larger and the asymmetry becomes smaller. However, the overall behaviors
are not too different from the $\theta =180^{\circ }$ case.

\begin{figure}
\includegraphics[  scale=0.5,
  angle=270,
  origin=lB]{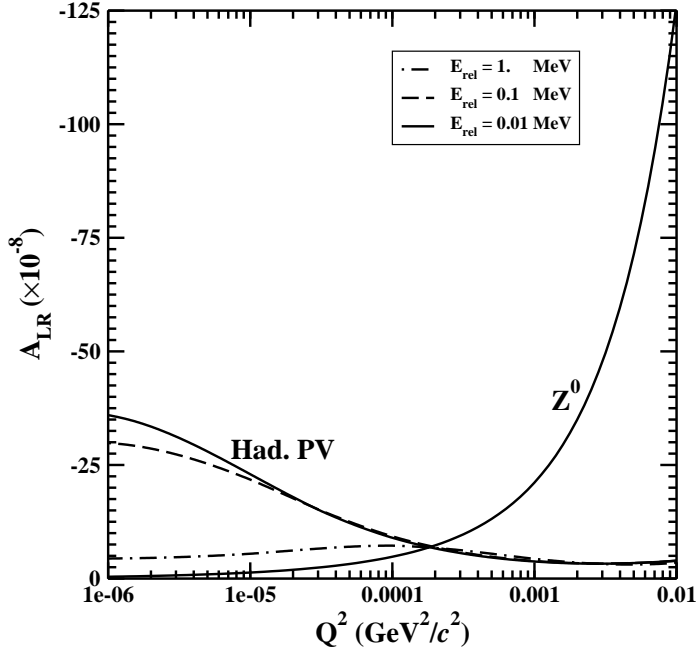}

\caption{The asymmetries due to hadronic PV and $Z^{0}$exchange versus $Q^{2}$
in the threshold electro-disintegration region, with small, fixed
$np$ relative energies. \label{fig:THD}}
\end{figure}

\begin{figure}
\includegraphics[  scale=0.5,
  angle=270,
  origin=lB]{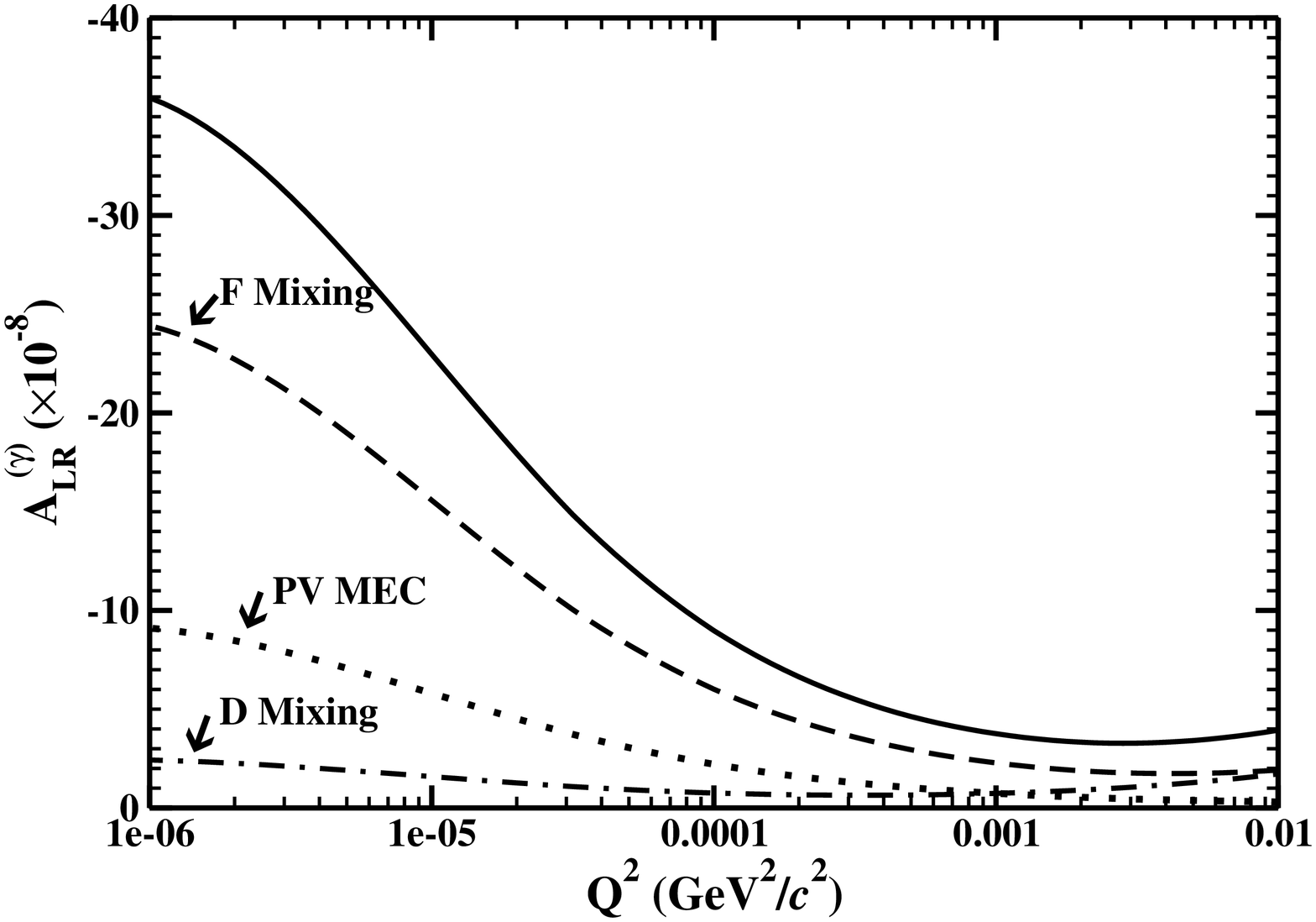}

\caption{The breakdown of various hadronic PV contributions to the asymmetry
in the threshold electro-disintegration region, where D, F, and MEC
refer to contributions from deuteron mixing, final state mixing, and
PV meson exchange currents, respectively, and the solid line gives
the total. \label{fig:breakdown THD}}
\end{figure}

Summarizing these observations, we conclude that the two-body hadronic
PV effects in QE $e$-$d$ scattering are negligible. However, the
situation changes in the kinematic region of threshold disintegration
as shown in Fig. \ref{fig:THD}. At $Q^{2}\sim 10^{-4}$(GeV/$c$)$^{2}$,
these two are comparable%
\footnote{The $Z^{0}$ asymmetry is plotted using the same formula from the
static approximation. Although there are also two-body effects, the
$Q^{2}$-dependence still governs the overall behavior.%
}, and hadronic PV dominates when moving toward lower $Q^{2}$ region.
Here again, the magnitude of $Q^2$ at which the hadronic PV and $Z^0$-exchange
contributions are commensurate is roughly what one would expect based on
the simple scaling arguments of Eq. (\ref{eq:scaling}).
The detailed breakdown given in Fig. \ref{fig:breakdown THD} shows that
the final state mixing has the most important contribution and that
PV meson exchange currents are also significant. The deuteron mixing,
still rather independent of $Q^{2}$ evolution, becomes negligible.
We also point out that while our calculation in this kinematic region
is consistent with Hwang $\etal $'s \cite{HwHM81} at $Q^{2}\ge 0.0001$
GeV$^{2}$/$c^{2}$ , we obtain larger asymmetries as one approaches
the threshold region. The reason is that we use a potential (AV$_{18}$)
which has a much softer core than the Reid soft-core potential. Thus,
the behavior of the wave function at low energy and small distance
is important for studies of hadronic PV at threshold, including experiments
like the photo-disintegration of deuteron, radiative neutron capture,
and neutron spin rotation.

\section{Conclusions \label{sec:summary}}

The theoretical analysis of PV electron scattering asymmetries requires that 
one
take into account effects which may, in principle, cloud the intended 
interpretation
of an experimental result. In this study, we have analyzed the effects of 
parity violating
NN interactions which give rise to a non-vanishing inelastic  $eD$ asymmetry 
at the
photon point. Our results indicate that for the QE kinematics relevant to the 
SAMPLE
experiment, these effects generate a negligible contribution to the PV 
asymmetry. Moreover,
contributions arising from each side of the QE peak produce cancellations 
when integrated
over detector acceptances, thereby generating an additional suppression of 
the nuclear
PV contamination. From this standpoint, then, the PV QE asymmetry provides a 
theoretically
clean environment for studying electroweak nucleon form factors, such as 
$\GAeV (Q^{2})$.
On the other hand, PV effects in the threshold region can become dominant, 
with asymmetries
as large as a few $\times 0.1$ ppm. Hence, near-threshold electro- or
photo-disintegration of the deuteron could provide a tool for probing the PV 
NN interaction.

\begin{acknowledgments}
We thank  T. Ito,  R.D. McKeown, T.W. Donnelly, and B. Jennings
for useful discussions; J. Carlson, M. Paris, and R. 
Schiavilla for making available to us
-- prior to publication -- the results of their analogous computation of 
nuclear PV contributions to the QE asymmetry and for comments on our
analysis; and the hospitality of Institute for Nuclear
Theory where part of this work has been accomplished. C.-P.L. would also 
like to thank G.A. Miller for useful discussions
on obtaining the wave functions. This work was support in part under DOE 
contract Nos.
DE-FG03-02ER41215 and DE-FG03-00ER41132, and NSF grant No. PHY-0071856.
\end{acknowledgments}
\bibliographystyle{apsrev}
\bibliography{QEPV}

\end{document}